\pdfoutput=1
\documentclass{aa2}   

% ...

\usepackage{txfonts} 
\usepackage{amsmath, amssymb}
\usepackage[english=british]{csquotes}
\usepackage[T1]{fontenc}
\usepackage[utf8]{inputenc}
\usepackage{natbib}
\usepackage{lmodern}
\usepackage{graphicx,rotating}
\graphicspath{{.}{plots/}}
\usepackage{calc}
 \bibpunct{(}{)}{;}{a}{}{,}
\usepackage{hyperref}
\usepackage[textwidth=3.5cm]{todonotes}
\usepackage{verbatim}
\usepackage{physics} 
\usepackage[alsoload=astro]{siunitx} 
\usepackage{multirow}
\usepackage{tabularx}
\usepackage{tikz-cd}

\usepackage{caption}
\usepackage{subcaption}

\usepackage{pstricks,pst-node}

\newcommand{\Omm}{\Omega_\mathrm{m}}
\newcommand{\Map}{\mathcal{M}_\mathrm{ap}}
\newcommand{\Mperp}{\mathcal{M}_{\perp}}
\newcommand{\MapMapMap}{{\expval{\Map^3}}}
\newcommand{\MapMperpMperp}{{\expval{\Map\Mperp^2}}}

\newcommand{\diracd}{\delta_\mathrm{D}}
\newcommand{\LCDM}{$\Lambda$CDM}
\newcommand{\bkappa}{\ensuremath{B_\kappa}}
\newcommand{\pkappa}{\ensuremath{P_\kappa}}
\newcommand{\MapEst}{\widehat{\mathcal{M}}_\mathrm{ap}}
\newcommand{\MperpEst}{\widehat{\mathcal{M}}_\perp}

\newcommand{\vtheta}{\vec{\theta}}
\newcommand{\vell}{\vec{\ell}}
\newcommand{\Npix}{N_\mathrm{pix}}
\newcommand{\Ngal}{N_\mathrm{gal}}

\newcommand{\gammao}{\gamma^{(\mathrm{o})}}

\newcommand{\myexpval}[1]{{\expval{#1}}}

\newcommand{\ee}{\mathrm{e}}
\newcommand{\ii}{\mathrm{i}}

\newcommand{\gammat}{\gamma_\mathrm{t}}
\newcommand{\gammax}{\gamma_\times}

\newcommand{\astroang}[1]{\ang[angle-symbol-over-decimal]{#1}}

\makeatletter
\renewcommand*\aa@pageof{page \thepage{} of \pageref*{LastPage}}
\makeatother

\hypersetup{pdfauthor={Name}}

\begin{document}

\title{
A roadmap to cosmological parameter analysis with third-order shear statistics I: Modelling and validation
}
\titlerunning{Cosmological parameter analysis with third-order shear statistics}

\author{Sven Heydenreich \inst{1},
        Laila Linke \inst{1},
        Pierre Burger \inst{1},
        Peter Schneider \inst{1}
      }
\authorrunning{Heydenreich et al.}

\institute{
      Argelander-Institut f\"ur Astronomie, Auf dem H\"ugel 71, 53121 Bonn, Germany 
      \\ \email{sven@astro.uni-bonn.de}
      }

\date{Version \today; received xxx, accepted yyy} 

\abstract
{
    In this work, which is the first of a series to prepare a cosmological parameter analysis with third-order cosmic shear statistics, we model both the shear three-point correlation functions $\Gamma^{(i)}$ and the third-order aperture statistics $\MapMapMap$ from the \textsc{BiHalofit} bispectrum model and validate these statistics with a series of N-body simulations.
    We then investigate how to bin the shear three-point correlation functions to achieve an unbiased estimate for third-order aperture statistics in real data.
    
    Finally, we perform a cosmological parameter analysis on KiDS1000-like mock data with second- and third-order statistics. We recover all cosmological parameters with very little bias. Furthermore, we find that a joint analysis almost doubles the constraining power on $S_8$ and increases the figure-of-merit in the $\Omm$-$\sigma_8$ plane by a factor of 5.9 with respect to an analysis with only second-order shear statistics.
    
    Our modelling pipeline is publicly available at \url{https://github.com/sheydenreich/threepoint/releases/}.
}


\maketitle

\section{Introduction}
\label{sec:introduction}

    The $\Lambda$ Cold Dark Matter model ($\Lambda$CDM) has been considered the standard model of cosmology for the past few decades. This relatively simple, 6-parameter model describes a wide range of observations, from the cosmic microwave background (CMB) to the observed large-scale structure of galaxies (LSS), with remarkable accuracy. As the reported uncertainties on cosmological parameters are approaching the per-cent level, a few tensions arise between CMB observations of the early Universe and observations of the local Universe that quantify the LSS \citep[for example, in the Hubble parameter $H_0$, see][and references therein]{diValentino:2021h}. In the past few years, also the matter clustering parameter $S_8=\sigma_8\sqrt{\Omm/0.3}$ has become subject to tension \citep[][and references therein]{Hildebrandt:2017, planck2020, joudaki2020, heymans2021,DES2021,diValentino:2021s}: The local Universe seems less clustered than observations of the CMB suggest. Assuming that these tensions are not due to unknown systematic effects, extensions to the $\Lambda$CDM model need to be explored. One of the most popular extensions is the $w$CDM model, where the equation-of-state of dark energy differs from $w=-1$.
    
    The dark energy task force has established that the weak gravitational lensing effect from the LSS, also called cosmic shear, poses one of the most promising methods to constrain the equation-of-state of dark energy \citep{Albrecht:2006}. The next generation of cosmic shear surveys like {\it Euclid} \citep{Laureijs:2011} or the Vera Rubin Observatory Legacy Survey of Space and Time \citep[LSST,][]{Ivezic:2008} will be able to constrain potential extensions to the $\Lambda$CDM model and may help to decipher the nature of dark energy.

    Tight constraints on cosmological parameters are essential to discriminate between the different modifications of the $\Lambda$CDM model. So far, two-point statistics have been established as the main analysis tool for cosmic shear \citep{Schneider:1998,Troxel:2018,Hildebrandt:2017,Hikage:2019,Asgari:2020,Hildebrandt:2020}. These statistics capture the entire information content of a Gaussian random field. Since the initial density field of the Universe is believed to be Gaussian, two-point statistics capture a large amount of cosmological information. However, in late times non-linear structure formation has introduced non-Gaussian features at the smaller scales of the matter distribution, whose information content cannot be captured by two-point statistics. To use this information, a variety of higher-order statistics has been introduced in recent years, including peak count statistics \citep{Martinet:2018,Harnois-Deraps:2021}, persistent homology \citep{Heydenreich:2021}, density split statistics \citep{Gruen:2018,Burger2022} and many others. In this work, we consider third-order shear statistics, which measure the skewness of the LSS at various scales.

    In contrast to most of these higher-order statistics, three-point statistics can be directly modelled from a matter bispectrum, allowing for a broad range of consistency checks that can be performed. Furthermore, their modelling does not require simulations that are adjusted to specific survey properties, which allows us to apply them to any data set easily.
    However, these natural extensions to two-point statistics have yet received surprisingly little attention.
    Several papers have reported a massive potential information gain when combining two- and three-point statistics \citep{Kilbinger:2005,Sato:2013,Kayo:2013}. \citet{Fu:2014} performed a combined analysis of two- and three-point statistics on  $154\,\mathrm{deg}^2$ CFHTLenS data, reporting a rather moderate gain in information content. \citet{Secco:2022} measured the shear three-point correlation functions and third-order aperture statistics in the third-year data release of the Dark Energy Survey \citep{Flaugher:2005,Sevilla-Noarbe:2021}, showing that they can be detected with high signal-to-noise. Recently, \citet{Pyne:2021} showed that a combined analysis of two- and three-point statistics has an additional advantage: These two statistics react differently to observational and astrophysical systematics, meaning that a combined analysis allows us to constrain nuisance parameters internally without the need for any additional observations or simulations, yielding an additional null-test and much tighter bounds on cosmological parameters (in an optimistic case, a factor of 20 in the figure-of-merit of dark energy can be achieved). 
    
    This article aims at preparing a cosmological parameter analysis with cosmic shear three-point statistics by developing a pipeline to measure and model both the three-point correlation functions $\Gamma^{(i)}$ and third-order aperture mass statistics $\MapMapMap$. We compare different estimators, computational costs and information content of both statistics. The covariance calculation will be investigated in the second publication of this series (Linke et al., in prep.). We further show that these statistics can be measured with relative ease in a Stage-III survey, which constitutes a significant advantage over their Fourier-space counterpart, the convergence bispectrum. The aperture mass statistics have several additional advantages: They offer good data compression, are not subject to the mass-sheet degeneracy, and, most importantly, they decompose the signal into E- and B-modes, where to leading order only E-modes can be created by gravitational lensing.
    
\begin{figure}
    \centering
    \begin{tikzcd}
         & & \Gamma^{(i)} \arrow{dd}[rotate=-90,yshift=1ex,anchor=center]{\footnotesize{\text{Sect.~\ref{subsec:measuring_map_from_3pcf}}}} & \\
        \Omega_\mathrm{m},\sigma_8,\ldots \arrow{r}{{{\text{Sect.~\ref{subsec:modelling_bispectrum}}}}} & B_\kappa \arrow{ru}[rotate=32,yshift=1ex,anchor=center]{\footnotesize{\text{Sect.~\ref{subsec:modelling_shear_3pcf}}}}\arrow{rd}[rotate=-35,yshift=1ex,anchor=center]{\footnotesize{\text{Sect.~\ref{subsec:modelling_map}}}} & &\arrow{lu}[rotate=-24,yshift=1ex,anchor=center]{\footnotesize{\text{Sect.~\ref{subsec:measuring_3pcf}}}}\arrow{ld}[rotate=26,yshift=1ex,anchor=center]{\footnotesize{\text{Sect.~\ref{subsec:measuring_map_direct}}}} \arrow[bend right=75]{ll}[yshift=2ex]{\footnotesize{\text{Sect.~\ref{subsec:measuring_bispectrum}}}}\text{Simulations} \\
        & & \left<\mathcal{M}_\mathrm{ap}^3\right> &
    \end{tikzcd}
    \caption{A schematic diagram of the modelling and validation pipeline introduced in this paper. The numbers on the arrows correspond to the respective section where this part of the pipeline is discussed.}
    \label{fig:diagram_analysis_pipeline}
    \end{figure}
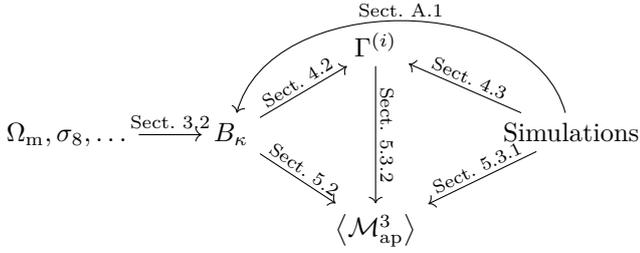

    In total, our modelling and validation pipeline can be summarised in the diagram in Fig.~\ref{fig:diagram_analysis_pipeline}. Our modelling algorithm is publicly available under \url{https://github.com/sheydenreich/threepoint/releases}.

    The paper is structured as follows: In Sect.~\ref{sec:nbody_sims} we introduce the N-body simulations we use to validate and test our modelling pipeline. We then present the convergence bispectrum in Sect.~\ref{sec:power_and_bispectra}, the shear three-point correlation functions in Sect.~\ref{sec:shear_3pcf} and the third-order aperture statistics in Sect.~\ref{sec:map3}. For each of these statistics, we describe their theoretical background, how we chose to model them, how they are measured in simulations, and the validation tests we performed. We then compare their information content to the one of second-order shear statistics in a mock-MCMC in Sect.~\ref{sec:results_mcmc} and discuss our findings in Sect.~\ref{sec:discussion}.

\section{Model validation and covariance estimation with $N$-body simulations}
    \label{sec:nbody_sims}
    We use N-body simulations containing only dark matter to validate our model and estimate covariance matrices. One of the main advantages of third-order shear statistics is that they can be easily adapted to different survey specifications. To highlight this, we use different simulation suites with varying source redshift distributions, galaxy number densities and cosmologies for the validation and parameter estimation. In this paper, we use the full-sky gravitational lensing simulations described in \citet[][hereafter T17]{Takahashi2017}, the Millennium Simulations \citep[][hereafter MS]{Springel:2005,Hilbert:2008}, and the  Scinet-LIghtCone Simulations \citep[][hereafter SLICS]{Harnois-Deraps:2018}.

\subsection{T17 simulations}
\label{sec:T17_description}

The T17 are used in this work to perform a realistic analysis of a survey that mimics the KiDS-1000 data and are constructed from a series of nested cubic boxes with side lengths of $L,2L,3L...$ placed around a fixed vertex representing the observer’s position, with $L=450\,\mathrm{Mpc}/h$. Each box is replicated eight times and placed around the observer using periodic boundary conditions. With the $N$-body code \textsc{gadget2} \citep{Springel2001} the gravitational evolution of $2048^3$ dark matter particles is simulated. Within each box, three spherical lens shells are constructed, each with a width of $150\,\mathrm{Mpc}/h$, which are then used by the public code \textsc{GRayTrix}\footnote{\url{http://th.nao.ac.jp/MEMBER/hamanatk/GRayTrix/}} to trace the light-ray trajectories from the observer to the last scattering surface\footnote{These maps are freely available for download at \url{http://cosmo.phys.hirosaki-u.ac.jp/takahasi/allsky_raytracing/}}. The cosmological parameters of the simulation are $\Omega_{\rm m}=1-\Omega_\Lambda=0.279$, $\Omega_{\rm b}=0.046$, $h=0.7$, $\sigma_8=0.82$, and $n_{\rm s}=0.97$. The matter power spectrum agrees with theoretical predictions of the revised \textsc{Halofit} \citep{Takahashi2012} within $5\%(10\%)$ for $k<5 (6)\,h\,\mathrm{Mpc}^{-1}$ at $z<1$.
\begin{figure}
\includegraphics[width=\columnwidth]{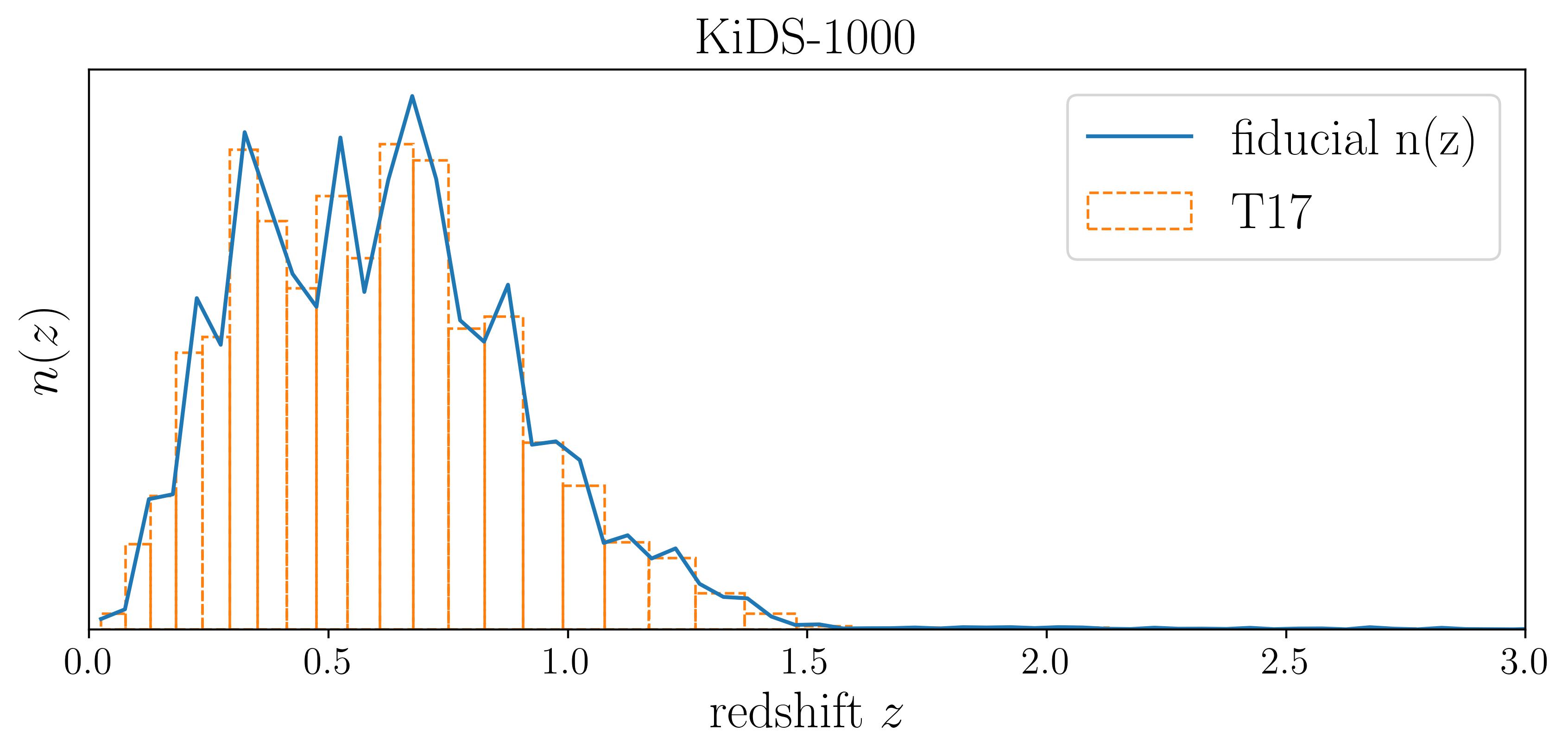}
\caption{Redshift distribution constructed from the T17 simulation given the fiducial $n(z)$ of the KiDS-1000 data.}
\label{fig:nofz}
\end{figure}
For each of the 108 realisations, we build a realistic convergence map by taking a weighted average of all 38 convergence shells at different redshifts, where the weights were determined by the fiducial KiDS-1000 $n(z)$ -- see Fig.~\ref{fig:nofz}.

We then transform the pure convergence maps into realistic convergence maps by adding to each pixel a Gaussian random variable with a vanishing mean and standard deviation of
\begin{equation}
    \sigma_\mathrm{pix} = \frac{\sigma_\epsilon}{\sqrt{n_\mathrm{gal}A_\mathrm{pix}}} \,,
\end{equation}
where $A_\mathrm{pix}$ is the pixel area of the convergence grid, and the effective number density $n_\mathrm{gal}=6.17\,\mathrm{arcmin}^{-2}$ and $\sigma_\epsilon=0.265$ are chosen such that they are consistent with the combined 1-5 tomographic bin of the KiDS-1000 data.

\subsection{Millennium Simulations}
    The MS were run with $2160^3$ particles in a $500\,h^{-1}\,\mathrm{Mpc}$ box in a flat {\LCDM} cosmology with $h=0.73$, $\sigma_8=0.9$, $\Omm=0.25$, $\Omega_\mathrm{b}=0.045$ and $n_s=1$. Subsequently, shear- and convergence-maps of 64 independent lines of sight with an area of $4\times 4\,\text{deg}^2$ each were created at 36 different redshifts \citep{Hilbert:2008,Hilbert:2009}. Each map is calculated on a grid of $4096\times 4096$ pixel. For each line of sight, we use the full shear- and convergence maps at redshift $z=1$. As we use the MS solely to validate our model, we do not add any noise to the maps.

\subsection{Scinet-LIghtCone Simulations}
    The SLICS were run with $1536^3$ particles in a $505\,h^{-1}\,\mathrm{Mpc}$ box, filling up $10\times 10\,\mathrm{deg}^2$ light-cones up to $z=3$. All SLICS were run in a flat {\LCDM} cosmology with $h=0.69$, $\sigma_8=0.83$, $\Omm=0.29$, $\Omega_\mathrm{b}=0.047$ and $n_s=0.969$. From these simulations, we use convergence maps as well as galaxy catalogues 
    with a redshift distribution of
    \begin{equation}
          n(z)\propto z^2\exp\left[-\left(\frac{z}{z_0}\right)^\beta\right]\; ,
    \end{equation}
    with $z_0=0.637$, $\beta=1.5$ and the overall proportionality constant given by normalising the distribution to $30\,\text{gal}/\text{arcmin}^2$. We use mock galaxy catalogues provided for 924 (pseudo-)independent lines of sight. The SLICS are useful to estimate the constraining power of third-order statistics since the three-point correlation function can be calculated relatively quickly on the $100\,\mathrm{deg}^2$ fields, and the 924 lines of sight enable the determination of a stable covariance matrix.
    
\section{Convergence power- and bispectrum}
    \label{sec:power_and_bispectra}
    In this section, we will briefly recap the basics of the weak gravitational lensing formalism, focusing on the shear statistics in Fourier space. More detailed reviews can be found in \citet{Bartelmann:2001,Hoekstra:2008, Munshi:2008, Bartelmann:2010}.
    
    We start by defining the density contrast at comoving position $\vec{x}$ and redshift $z$, $\delta(\vec{x},z) = \frac{\rho(\vec{x},z)}{\bar{\rho}(z)}-1$, where $\rho(\vec{x},z)$ is the matter density at position $\vec{x}$ and redshift $z$ and $\bar{\rho}(z)$ the average density at redshift $z$.
    
    From this density contrast, we define the convergence $\kappa$ for sources  at redshift $z$ as a line-of-sight integration, weighted by the lensing efficiency
    \begin{align}
        \kappa(\vec{\theta},z) = \frac{3\Omm H_0^2}{2 c^2}\int_0^{\chi(z)}\dd \chi'{}&{}\, \frac{f_K[\chi'(z)]\,f_K[\chi'-\chi(z)]}{f_K[\chi(z)]}\nonumber\\
        {}&{}\times\frac{\delta[f_K(\chi')\vec{\theta},z]}{a(\chi')}\; ,
    \end{align}
    where $f_K(\chi)$ is the comoving angular diameter distance. We note that throughout this paper, we work in a flat Universe with $\Omm+\Omega_\Lambda=1$, meaning that $f_K[\chi(z)]=\chi(z)$, where $\chi(z)$ is the comoving distance at redshift $z$. However, everything we present in this section also works for open or closed universes. The convergence can not be directly observed, but it can be recovered from an observed shear field \citep{Kaiser:1995, Seitz:1998, Jeffrey:2020}. The relations between shear, convergence and the matter density contrast allow us to relate all second- and third-order shear statistics to the well-understood matter power spectrum $P_\delta(k,z)$ and bispectrum $B_\delta(k_1,k_2,k_3,z)$.
    \subsection{Definition of power- and bispectrum}
        \label{subsec:power_and_bispectra_theory}
        The matter power spectrum $P_\delta(k,z)$ and bispectrum $B_\delta(k_1,k_2,k_3,z)$ can be defined as 
        \begin{align}
            \expval{\hat{\delta}(\vec{k}_1,z)\hat{\delta}(\vec{k}_2,z)} = {}&{} (2\pi)^3\,\delta_\mathrm{D}(\vec{k}_1+\vec{k}_2)\,P_\delta(k_1,z) \; , \\
            \expval{\hat{\delta}(\vec{k}_1,z)\hat{\delta}(\vec{k}_2,z)\hat{\delta}(\vec{k}_3,z)} = {}&{}  (2\pi)^3\,\delta_\mathrm{D}(\vec{k}_1+\vec{k}_2+\vec{k}_3) \nonumber\\
            {}&{}\qquad \times B_\delta(k_1,k_2,k_3,z) \; , 
            \label{eq:defn_bispectrum}
        \end{align}
        where $\hat{\delta}$ describes the Fourier transform of $\delta$ and $\delta_{\rm D}$ is the Dirac-delta distribution. The fact that the power- and bispectrum only depend on the moduli of the $k$-vectors can be easily derived from the statistical isotropy of the Universe.

        The convergence power- and bispectrum can then be computed using the Limber approximation \citep{Limber:1954,Peebles:1980,Kaiser:1997,Bernardeau:1997,Schneider:1998},
        \begin{align}
            \pkappa (\ell) = {}&{} \frac{9\Omm^2H_0^4}{4c^4}\int_0^{\chi_\mathrm{max}} \dd \chi\;\frac{g^2(\chi)}{a^2(\chi)} \nonumber \\ {}&{}\qquad\times P\left[\frac{\ell}{f_K(\chi)},z(\chi)\right] \;, \label{eq:pkappa_defn}\\
            \bkappa(\ell_1,\ell_2,\ell_3) = {}&{} \frac{27H_0^6\Omm^3}{8c^6}\int_0^{\chi_\mathrm{max}}\dd \chi\; \frac{g^3(\chi)}{a^3(\chi)\,f_K(\chi)} \nonumber \\ {}&{}\times B_\delta\left[\frac{\ell_1}{f_K(\chi)},\frac{\ell_2}{f_K(\chi)},\frac{\ell_3}{f_K(\chi)},z(\chi)\right] \label{eq:bkappa_defn}\, .
        \end{align}
        Here,
        \begin{equation}
            g(\chi) = \int_\chi^{\chi_\mathrm{max}} \dd \chi'\; p(\chi')\,\frac{f_K(\chi'-\chi)}{f_K(\chi')} 
            \label{eq:lensing_efficiency_defn}
        \end{equation}
        describes the lensing efficiency, where $p(\chi)$ is the (normalised) comoving distance probability distribution of sources. We note that usually one instead measures a redshift probability distribution $p(z)$; in our modelling pipeline we thus instead write Eqs.~\eqref{eq:pkappa_defn}, \eqref{eq:bkappa_defn} and \eqref{eq:lensing_efficiency_defn} as integrals over the redshift $z$.

        The Limber approximation breaks down for small values of $\ell$. For example, the bispectrum from the Limber approximation overestimates the truth by up to an order of magnitude for $\ell\ll 60$ (corresponding to angular scales of roughly \astroang{6;;}), depending on the source redshift distribution \citep{Deshpande:2020}. However, at these scales, the impact of non-linear structure formation is small, meaning that the matter distribution is well-described by a Gaussian and higher-order statistics like the bispectrum are small. In this work, we only consider shear statistics up to $\sim\astroang{4;;}$; at these scales, we expect the Limber approximation to hold. 

        Instead of the three $\ell$-values $\ell_1,\ell_2$ and $\ell_3$, we can also describe the bispectrum as a function of the two vectors $\vec{\ell}_1,\vec{\ell}_2$ or their moduli $\ell_1,\ell_2$ and the angle $\varphi$ between them. We define
        \begin{equation}
            b(\ell_1,\ell_2,\varphi)=B_\kappa\left(\ell_1,\ell_2,\sqrt{\ell_1^2+\ell_2^2+2\ell_1\ell_2\cos\varphi}\right)\; .
        \end{equation}
        
    \subsection{Modelling the bispectrum}
        \label{subsec:modelling_bispectrum}
        We use the state-of-the-art \textsc{BiHalofit} algorithm \citep{Takahashi:2020} to model the dark matter bispectrum on non-linear scales. In comparison to older bispectrum models \citep[e.g.][]{Gil-Marin:2012,Scoccimarro:2001}, \textsc{BiHalofit} appears to trace the non-equilateral triangles much better: In comparison with N-body simulations, \textsc{BiHalofit} retains an accuracy of $10\%$ or better, whereas the other two fitting formulae are subject to errors of more than $200\%$. The effects on higher-order shear statistics are substantial, as can be seen in \citet{Halder:2021}, who modelled a different third-order shear statistic from using both  \textsc{BiHalofit} and the bispectrum model of \citet{Gil-Marin:2012}. In a direct comparison with N-body simulations, \textsc{BiHalofit} is accurate on all tested scales, whereas the other fitting formula breaks down on scales of $\lesssim\astroang{;30;}$.
        
        Another advantage of \textsc{BiHalofit} is that it only requires a linear power spectrum as its input, compared to a non-linear spectrum in \citet{Gil-Marin:2012} or \citet{Scoccimarro:2001}. We use the fitting formula developed by \citet{Eisenstein:1999} to model the linear power spectrum.
        
        One of the main advantages of third-order shear statistics is that one can rigorously test each stage of the modelling pipeline. We perform such a test on our bispectrum model in App.~\ref{sec:app_testing_bispectrum} and conclude that the Limber integrated \textsc{BiHalofit} bispectrum is consistent with the MS up to $\ell\lesssim 10^4$ and deviates by up to 40\% for larger values $\ell$.

\section{Shear three-point correlation functions}
    \label{sec:shear_3pcf}
    \subsection{Definition of the shear three-point correlation functions}
        \label{subsec:definition_shear_3pcf}
        Shear three-point correlation functions (3pcf) are the natural extension to the widely used two-point correlation functions. Let $\gamma_\mathrm{c}=\gamma_1+\mathrm{i}\gamma_2$ denote the complex shear in Cartesian coordinates.
        Considering a triangle of galaxies, as a first step, we project the shear $\gamma^i$ of each galaxy $i$ to its tangential- and cross-components with respect to a point fixed with respect to the triangle, for example one of its centres,
        \begin{equation}
            \label{eq:gamma_t_defn}
          \gamma\equiv\gamma_\mathrm{t} + \ii\gamma_\times = -\gamma_\mathrm{c}\ee^{-2\ii\zeta}\; ,
        \end{equation}
        where $\zeta$ is the angle of the projection direction. \citet{Schneider:2003} established four \emph{natural components} of the shear 3pcf, which remain invariant under rotations of the triangle. They are defined as
        \begin{figure*}
            \centering
            \sidecaption
            \def\svgwidth{12cm}
            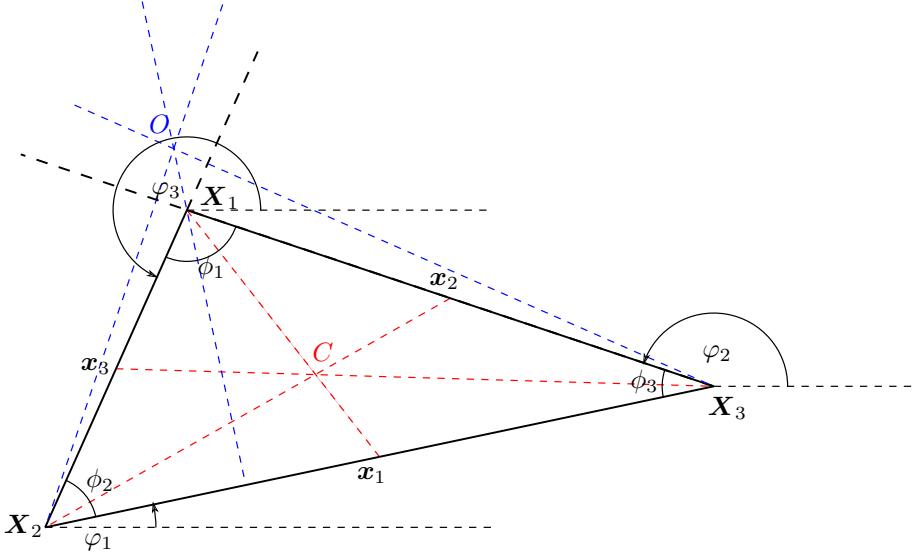
            \caption{An example triangle with our used notations. The sidelengths are denoted by $\vec{x}_1=\vec{X}_3-\vec{X}_2$ and cyclic permutations. The interior angles of the triangle are denoted by $\phi_i$, whereas $\varphi_i$ are the angles that the sidelengths take with respect to the $x$-axis. We also show two possible definitions of its centre: The orthocenter $O$ is at the intersection of the three altitudes (blue dashed); the centroid $C$ is at the intersection of its three medians (red dashed). Figure adapted from \citet{Schneider:2005}.}
            \label{fig:triangle_centers}
        \end{figure*}
        \begin{align}
        \label{eq:defn_natural_components}
            \Gamma^{(0)}=\expval{\gamma\gamma\gamma}\, {}&{}, \, \Gamma^{(1)}=\expval{\gamma^*\gamma\gamma}\, , \nonumber\\ \Gamma^{(2)}=\expval{\gamma\gamma^*\gamma}\, {}&{}, \, \Gamma^{(3)}=\expval{\gamma\gamma\gamma^*}\, ,
          \end{align}
        where the `$^*$' denotes complex conjugation.
        The choice of the reference point of the projection is to some degree arbitrary. Usually one of the triangles cenres is chosen, most often the orthocenter (the intersection of its three altitudes) or the centroid (the intersection of its three medians), as shown in Fig.~\ref{fig:triangle_centers}. The natural components have the advantage that they are invariant under the choice of triangle centre up to multiplication with a complex phase factor, meaning that their moduli are invariant under the choice of triangle centre.

        Parametrizing the shear 3pcf by the triangle side-lengths, $x_1, x_2$ and $x_3$, where the indices 1,2 and 3 are ordered in a counter-clockwise direction, the natural components exhibit a nice behaviour concerning cyclic permutation of arguments. While the first natural component $\Gamma^{(0)}$ is invariant under cyclic permutations, the other three components transform into each other,
        \begin{align}
        \Gamma^{(0)}(x_1,x_2,x_3) = {}
        &\Gamma^{(0)}(x_2,x_3,x_1)=\Gamma^{(0)}(x_3,x_1,x_2) 
        \;,\nonumber \\
        \Gamma^{(1)}(x_1,x_2,x_3) = {}
        &\Gamma^{(3)}(x_2,x_3,x_1)=\Gamma^{(2)}(x_3,x_1,x_2) \; .
        \label{eq:permutations_natural_components}
        \end{align}
        Similar behaviour can be observed for parity transformations:
        \begin{align}
            \Gamma^{(0)}(x_1,x_2,x_3) = {}&\Gamma^{(0)*}(x_2,x_1,x_3)\;,\nonumber\\
            \Gamma^{(1)}(x_1,x_2,x_3) = {}&\Gamma^{(1)*}(x_1,x_3,x_2)\;,\nonumber\\
            \Gamma^{(2)}(x_1,x_2,x_3) = {}&\Gamma^{(2)*}(x_3,x_2,x_1)\;,\\
            \Gamma^{(3)}(x_1,x_2,x_3) = {}&\Gamma^{(3)*}(x_2,x_1,x_3)\;.\nonumber
        \end{align}

    \subsection{Modelling the shear three-point correlation functions}
        \label{subsec:modelling_shear_3pcf}
        We model the natural components $\left\{\Gamma^{(i)}\right\}_{i=0,1,2,3}$ of the three-point correlation functions of cosmic shear using the methods described in \citet[][hereafter S+05]{Schneider:2005}: 
        When we project the shear of all three galaxies to the orthocenter of the triangle, the projection direction at the triangle vertex $\vec{X}_i$ is orthogonal to the orientation $\varphi_i$ of the triangle side $\vec{x}_i$. Thus, the shear transforms as
        \begin{equation}
            \label{eq:rotation_orthocenter}
            \gamma^{(\mathrm{o})}(\vec{X}_i)=\gamma_\mathrm{c}(\vec{X}_i)\ee^{-2\ii\varphi_i}\; .
        \end{equation}
        We now utilise the relation between convergence and shear in Fourier space \citep{Kaiser:1993},
        \begin{equation}
            \hat{\gamma}_\mathrm{c}(\vec{\ell})=\ee^{2\ii\beta_\ell}\hat{\kappa}(\vec{\ell})\; ,
        \end{equation}
        where $\beta_\ell$ is the polar angle of $\vec{\ell}$, to write
        \begin{align}
			\Gamma^{(0)}{}&{}(x_1,x_2,x_3) = \myexpval{\gammao(\vec{X}_1)\gammao(\vec{X}_2)\gammao(\vec{X}_3)} \notag\\
			= {}&{} \int\frac{\dd^2\ell_1}{(2\pi)^2}\int\frac{\dd^2\ell_2}{(2\pi)^2}\int\frac{\dd^2\ell_3}{(2\pi)^2}\; \myexpval{\tilde{\kappa}(\vec{\ell}_1)\tilde{\kappa}(\vec{\ell}_2)\tilde{\kappa}(\vec{\ell}_3)} \notag\\
			{}&{}\times \exp\left[-\ii(\vec{\ell}_1\cdot\vec{X}_1+\vec{\ell}_2\cdot\vec{X}_2+\vec{\ell}_3\cdot\vec{X}_3)\right]\\
			{}&{}\times\exp\left[2\ii\sum_i\left(\beta_i-\varphi_i\right)\right] \; . \notag
        \end{align}
            This can then be transformed into equation (15) of S+05\footnote{We note that due to the different definitions of the convergence bispectrum (Eq.~\ref{eq:defn_bispectrum} vs equation 4 in S+05), we get a factor of 3 difference.}:
            \begin{align}
                  \Gamma^{(0)}{}&{}(x_1,x_2,x_3) = \frac{2\pi}{3}\int_0^\infty\frac{\dd\ell_1\,\ell_1}{(2\pi)^2}\int_0^\infty\frac{\dd\ell_2\,\ell_2}{(2\pi)^2}\int_0^{2\pi}\dd{\varphi} \nonumber\\
                  {}&{} \times b(\ell_1,\ell_2,\varphi)\, \ee^{2\ii\bar{\beta}} \left[\ee^{\ii(\phi_1-\phi_2-6\alpha_3)}J_6(A_3) \right. \\
                  {}&{} \left. + \ee^{\ii(\phi_3-\phi_2-6\alpha_1)}J_6(A_1) + \ee^{i(\phi_3-\phi_1-6\alpha_2)}J_6(A_2) \right] \; , \nonumber
            \end{align}
            with
            \begin{align}
                  &A_3 = \left[(\ell_1x_2)^2 + (\ell_2x_1)^2
                  + x_1x_2\ell_1\ell_2\cos(\varphi+\phi_3)\right]^{\frac{1}{2}} \nonumber\; , &&&\\
                  &|\vec{\ell}_1+\vec{\ell}_2|^2 \cos2\bar{\beta} = (\ell_1^2+\ell_2^2)\cos\varphi+2\ell_1\ell_2 \; ,&&&\nonumber\\
                  &|\vec{\ell}_1+\vec{\ell}_2|^2 \sin 2\bar{\beta} =  (\ell_1^2-\ell_2^2)\sin\varphi\; ,&&&\\
                  & A_3\cos\alpha_3 = (\ell_1x_2+\ell_2x_1)\cos\left(\frac{\varphi+\phi_3}{2}\right) \; ,&&&\nonumber\\
                    &A_3\sin\alpha_3 = (\ell_1x_2-\ell_2x_1)\sin\left(\frac{\varphi+\phi_3}{2}\right) \; .&&& \nonumber
            \end{align}
            The quantities $A_{1,2}$ and $\alpha_{1,2}$ are obtained by cyclic permutation of indices. The angles $\phi_i$ are the interior angles of the triangle, as shown in Fig.~\ref{fig:triangle_centers}.
            By introducing polar coordinates $R=\sqrt{\ell_1^2+\ell_2^2}$, $\psi = \arctan(\ell_2/\ell_1)$, we get
            \begin{align}
                  \Gamma^{(0)}{}&{}(x_1,x_2,x_3) =  \frac{1}{3(2\pi)^3}\int_0^{2\pi}\dd{\varphi} \int_0^{\pi/2}\dd{\psi} \int_0^\infty \dd{R}\,\nonumber\\
                  {}&{}\times R^3\sin\psi\cos\psi\, b(R\cos\psi,R\sin\psi,\varphi)\,\ee^{2\ii\bar{\beta}} \\
                  {}&{}\times  \left[\ee^{\ii(\phi_1-\phi_2-6\alpha_3)}J_6(R\,A_3') + \ee^{\ii(\phi_3-\phi_2-6\alpha_1)}J_6(R\,A_1') \right. \nonumber\\ {}&{} \quad + \left.\ee^{i(\phi_3-\phi_1-6\alpha_2)}J_6(R\,A_2') \right] \, , \nonumber
            \end{align}
            with
            \begin{align}
                  &A'_3 = \frac{A_3}{R} = \left[(x_2\cos\psi)^2 + (x_1\sin\psi)^2  \right.\nonumber\\
                  {}&{}\qquad\qquad\qquad\left.+ x_1x_2\sin2\psi\,\cos(\varphi+\phi_3)\right]^{\frac{1}{2}} \; ,  \nonumber&&&\\
                  &\cos2\bar{\beta} = (\cos\varphi+2\cos\psi\,\sin\psi) \; ,  \nonumber&&&\\
                  &\sin 2\bar{\beta} =  (\cos^2\psi-\sin^2\psi)\sin\varphi \; , &&&\\
                   &A'_3\cos\alpha_3 =  (x_2\cos\psi + x_1\sin\psi )\cos\left(\frac{\varphi+\phi_3}{2}\right) \; , &&&\nonumber\\
                    &A'_3\sin\alpha_3 =  (x_2\cos\psi-x_1\sin\psi )\sin\left(\frac{\varphi+\phi_3}{2}\right) \; . \nonumber &&&
            \end{align}
            Defining 
            \begin{equation}
                  E_3 = \ee^{\ii(\phi_1-\phi_2-6\alpha_3)} \; ,
            \end{equation}
            and $E_1$ and $E_2$ via cyclic permutations of indices, we can write
            \begin{align}
                  \Gamma^{(0)}(x_1,x_2,x_3) = {}&{} \frac{1}{6(2\pi)^5} \int_0^{\pi/2}\dd{\psi} \sin2\psi\int_0^{2\pi}\dd{\varphi} \nonumber \\ {}&{}\times \ee^{2\ii\bar{\beta}}\sum_{i=1}^3\frac{E_i}{A_i'^4}\int_0^\infty \dd{R}\,R^3 \label{eq:gamma0_from_bkappa}\\
                  & \times b\left(\frac{R}{A_i'}\cos(\psi),\frac{R}{A_i'}\sin(\psi),\varphi\right)J_6(R)\; . \nonumber
            \end{align}
        The $R$-integration filters the bispectrum with a 6-th order Bessel function, making the integration routine difficult to solve numerically, as the functional form of the bispectrum prevents the application of fast Hankel transform algorithms like FFTLog \citep{Hamilton:2000}. We thus use the method developed in \citet{Ogata:2005} to solve the $R$-integration and integrate the remaining dimensions using the \textsc{cubature} library.\footnote{\url{https://github.com/stevengj/cubature}}
        
        To model $\Gamma^{(1)}$, we apply the same transformations to equation (18) of S+05; for $\Gamma^{(2)}$ and $\Gamma^{(3)}$ we perform a cyclic permutation of the input variables as outlined in Eq.~(\ref{eq:permutations_natural_components}).
        
        While a triangle of galaxy positions for which we evaluate the three-point correlation function can be described by its three side-lengths $x_1, x_2$ and $x_3$, it is certainly not a good idea to use these variables for a binning scheme; for example, for $x_1>x_2+x_3$ a triangle can not be defined, which means that the 3pcf would not be defined for many bins. A better way to bin the triangles was introduced by \citet{Jarvis:2004}. Assuming $x_1>x_2>x_3$ they defined a triangle via the values $r\in[0,\infty]$, $u\in [0,1]$ and $v\in [-1,1]$ by
        \begin{equation}
            r = x_2,\quad u=\frac{x_3}{x_2},\quad v=\pm\frac{x_1-x_2}{x_3} \; .
        \end{equation}
        Here, $v$ is positive for triangles where $x_1,x_2$ and $x_3$ are oriented clockwise and negative for a counter-clockwise orientation. This binning choice allows us to bin the triangle size $r$ in logarithmic steps without having bins where the 3pcf is not defined. In all cases, we bin the shear 3pcf logarithmic in $r$ and linear in $u$ and $v$.
        
        We note that Eq.~\eqref{eq:permutations_natural_components} implies that the four shear 3pcf for $x_1>x_2>x_3$ \citep[as in the binning scheme of][]{Jarvis:2004} already contain the entire information content of the third-order shear signal, as does knowledge of $\Gamma^{(0)}$ and $\Gamma^{(1)}$ for all combinations of $x_1,x_2$ and $x_3$. In a similar manner, Eq~\eqref{eq:permutations_natural_components} implies that $\Gamma^{(i)}(r,u,v)=\Gamma^{(i)*}(r,u,-v)$ holds, where the `${}^*$' denotes complex conjugation.
        
        To ensure compatibility with results from the measured 3pcf, we transform the modelled functions from the centroid (as used in S+05) to the orthocenter \citep[as used in][compare Sect.~\ref{subsec:measuring_3pcf}]{Jarvis:2004}.
        
        For a potential cosmological parameter analysis, the three-point correlation functions face a few hurdles: Assuming we bin the three-point correlation functions in 10 bins for each $r,u$ and $v$, then our data vector consists of $8000$ entries. This makes estimating a covariance matrix using simulations practically impossible and leads to a modelling time that is unfeasible even for a non-tomographic analysis.
        
    \subsection{Measuring the shear three-point correlation functions}
        \label{subsec:measuring_3pcf}
        We use the public tree-code \textsc{treecorr} \citep{Jarvis:2004} to measure the three-point correlation functions $\Gamma_i$. This algorithm estimates the quantity
        \begin{equation}
            \widehat{\Gamma}^{(0)} = \frac{\sum_{ijk}w_i\varepsilon_{i}\,w_j\varepsilon_{j}\,w_k\varepsilon_{k}}{\sum_{ijk}w_iw_jw_k}\; ,
        \end{equation}
        where the $\varepsilon_i = \varepsilon_{\mathrm{t},i}+\mathrm{i}\varepsilon_{\times,i}$ are the observed ellipticities of galaxies and $w_i$ the associated weights. The other natural components $\widehat{\Gamma}^{(1)},\dots$ are estimated in the same manner. This estimator has the  advantage that it is not impacted by the survey geometry: As long as at least one galaxy triplet falls into each bin, it remains unbiased \citep{Simon:2008}.\footnote{Even if certain bins remain empty, the estimated correlation function can be rebinned with a tesselation scheme to yield unbiased values of $\Gamma^{(i)}$ for all bins, as was shown by \citet{Linke:2020} for the related galaxy-galaxy-galaxy-lensing correlation function.} The disadvantage of the estimator is that its computational complexity scales with $\mathcal{O}(\Ngal^3)$, which is not feasible to execute even for moderate values of $\Ngal\gtrsim 10^6$. That is why \textsc{treecorr} constructs a hierarchical ball tree out of the galaxy sample and calculates the correlation functions from this tree. This results in a remarkable speed-up and allows us to calculate the shear 3pcf for an ensemble of about $10^7$ source galaxies distributed over a $10\times 10\,\mathrm{deg}^2$ field in about $1\,500$ CPUh. A disadvantage of the tree-code is that its execution time scales massively with the number of bins: If $b$ is the logarithmic bin size, then the run-time scales roughly with $b^{-4}$.
        
        The \textsc{treecorr} algorithm also has a \textsc{binslop} parameter, which allows balls of the KD-tree to overlap the edges of a bin. This parameter heavily affects computation time, and while the expectation value is relatively stable between different values of \textsc{binslop}, the covariance is subject to change \citep{Secco:2022}.
        
    \subsection{Validation}
    We test our developed integration routine described in Sect.~\ref{subsec:modelling_shear_3pcf} with a lensing potential for which we can derive analytic expressions for both the convergence bispectrum and the shear 3pcf. As discussed in more detail in App.~\ref{sec:app_test_gamma_integration} we found an agreement to the sub-percent level.
        
    \label{subsec:validation_3pcf_vs_Nbody}
    \begin{figure*}
    \centering
    \includegraphics[width=17cm]{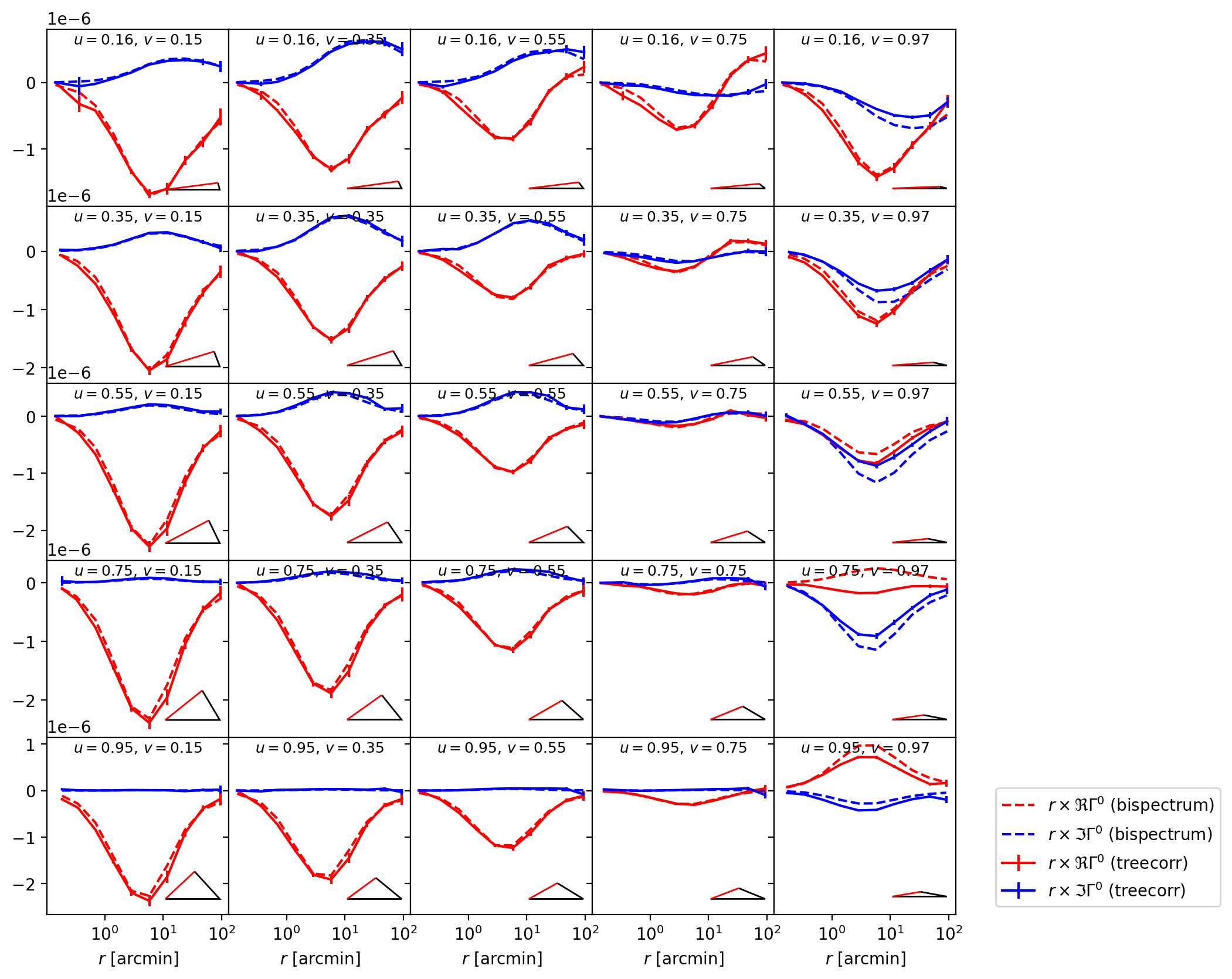}
    \caption{The first natural component $\Gamma^{(0)}$ of the shear three-point correlation functions modelled from \textsc{BiHalofit} compared to measurements from the MS that were extracted using \textsc{TreeCorr}. Each panel corresponds to one fixed triangle configuration; the $u$ and $v$ values are listed at the top, and the corresponding shape of the triangle is shown on the bottom right. The $x$-axis corresponds to the length of the red side of the triangle. We show both the real part (red) and the imaginary part (blue) for the model (dashed line) and the simulations (solid line). The error bars denote the error on the mean of the 64 lines of sight of the MS.}
    \label{fig:gamma0_bihalofit_vs_MS}
    \end{figure*}
    To validate our model for the shear 3pcf, we measure the shear signal at a redshift of $z=1$ in the MS. We choose to measure the signal in $10^3$ bins (10 bins in each $r,u$ and $v$), with logarithmic $r$-bins from $\astroang{;0.1;}$ to $\astroang{;120;}$. To speed up computation time, we randomly select every tenth pixel of the $4096^2$ pixel grid. As we do not include shape noise, we expect the loss of signal to be small. The results can be seen in Fig.~\ref{fig:gamma0_bihalofit_vs_MS}. We conclude that we can model the shear 3pcf reliably down to sub-arcminute scales for almost all triangle configurations. Only for almost degenerate, flattened triangle configurations ($v>0.9$) do we see that the model and simulations differ significantly. This might signify that \textsc{BiHalofit} breaks down at the corresponding triangle configurations in Fourier-space. Alternatively, this might point towards a break-down of the tree-code's accuracy at these very degenerate triangles. As we can see in Sect.~\ref{subsubsec:validation_n_body_sims_map}, these points play a negligible role in the conversion to aperture mass statistics. Overall we see that the agreement between shear 3pcf is better than the one at the bispectrum level (compare Sect.~\ref{subsubsec:validation_n_body_sims_bispectrum}).
            
    We observe the same effects for the other natural components of the 3pcf (compare Fig.~\ref{fig:gamma1_bihalofit_vs_MS}).

\section{Aperture mass statistics}
    \label{sec:map3}
    \subsection{Definition of aperture mass statistics} 
        \label{subsec:background_map3}
        An alternative way to analyze cosmic shear is via aperture mass maps \citep{Schneider:1996,Bartelmann:2001}. Their advantage is that they can separate the signal into so-called E- and B-modes \citep{Schneider:2002}, where B-modes can, to leading order, not be created by the weak gravitational lensing effect. Thus, the absence of B-modes provides a crucial null-test in a cosmic shear analysis \citep[see, e.g.,][]{Hildebrandt:2017,Asgari:2019}. Compared to convergence maps \citep{Kaiser:1993,Seitz:2001,Gatti:2021}, aperture mass maps are constructed in a way that they do not suffer from the well-known mass-sheet degeneracy \citep{Falco:1985,1995A&A...294..411S}.

        The aperture mass $\Map$ at position $\vec{\vartheta}$ and filter radius $\theta$ are defined as
        \begin{equation}
        \label{eq:definition_aperture_mass_kappa}
          \Map(\vec{\vartheta};\theta)=\int\dd^2\vartheta'\; U_\theta(|\vec{\vartheta}-\vec{\vartheta'}|)\, \kappa(\vec{\vartheta'}) \; ;
        \end{equation}
        here, $U_\theta(\vartheta)$ is a compensated filter (i.e. $\int \dd\vartheta\,\vartheta\, U(\vartheta) = 0$). Given a shear field $\gamma$, the aperture mass $\Map$ and its respective B-mode counterpart $\Mperp$ can be calculated as
        \begin{align}
            \label{eq:definition_aperture_mass_gamma}
              \Map(\vec{\vartheta};\theta)+\ii \Mperp(\vec{\vartheta};\theta)
              =  {}&{} \int\dd^2\vartheta' \; Q_\theta(|\vec{\vartheta}-\vec{\vartheta'}|)\nonumber\\ {}&{}\times\left[\gamma_\mathrm{t}(\vec{\vartheta'})+\ii\gammax(\vec{\vartheta'})\right] \; ,
        \end{align}
        where $\gammat$ and $\gammax$ are projected along the vector $\vec{\vartheta}'-\vec{\vartheta}$ (see Eq.~\ref{eq:gamma_t_defn})  and $Q_\theta$ is related to $U_\theta$ via
        \begin{equation}
          Q_\theta(\vartheta) = \frac{2}{\vartheta^2}\int_0^\vartheta \dd\vartheta'\;\vartheta'\, U_\theta(\vartheta')-U_\theta(\vartheta)\; .
        \end{equation}
        For simplicity of notation, we define $U_\theta(\vartheta)=\theta^{-2}u(\vartheta/\theta)$ and denote by $\hat{u}(\eta)$ the Fourier transform of $u$.
        In this work, we opt for the filter function introduced by \citet{Crittenden:2002},
        \begin{align}
          u(x)= {}&{}\frac{1}{2\pi}\left(1-\frac{x^2}{2}\right)\ee^{-x^2/2},\quad \hat{u}(\eta) = \frac{\eta^2}{2}\ee^{-\eta^2/2}, \nonumber\\
          Q_\theta(\vartheta) = {}&{} \frac{\vartheta^2}{4\pi\theta^4}\exp\left(-\frac{\vartheta^2}{2\theta^2}\right)\; .
        \end{align}

        While the construction of aperture mass maps has its uses \citep[see for example][]{Harnois-Deraps:2021,Heydenreich:2021}, we do not care about the structure of an aperture mass map but rather about its statistical properties. We define for arbitrary combinations of E- and B-mode aperture mass statistics
        \begin{align}
            {\expval{\Map^m\Mperp^n}}{}&{}(\theta_1,\ldots,\theta_n) =  \left<\Map(\vec{\vartheta};\theta_1)\dots\Map(\vec{\vartheta};\theta_m)\right.\nonumber\\
            {}&{}\times\left.\Mperp(\vec{\vartheta};\theta_m+1)\dots\Mperp(\vec{\vartheta};\theta_m+n)\right>_{\vec{\vartheta}} \; .    
        \end{align}
        By construction, $\expval{\Map}(\theta)$ vanishes. In a parity-symmetric field, all odd powers of B-mode components vanish \citep{Schneider:2003b}, meaning that the relevant B-mode counterparts to $\expval{\Map^2}$ and $\MapMapMap$ are $\expval{\Mperp^2}$ and $\MapMperpMperp$, respectively.

    \subsection{Modelling aperture mass statistics}
        \label{subsec:modelling_map}
        Given a model for the convergence power spectrum, the second-order aperture statistics can be calculated as
        \begin{equation}
            \expval{\Map^2}(\theta) = \int\frac{\dd \ell\;\ell}{2\pi}\,P_\kappa(\ell)\,\hat{u}^2(\theta\ell)\; .
        \end{equation}
        As a model for the non-linear power spectrum, we use the revised \textsc{Halofit} model of \citet{Takahashi2012}. Equivalently, the third-order aperture statistics $\MapMapMap$ can be derived from a bispectrum model via \citep[compare][]{Jarvis:2004,Schneider:2005}\footnote{Again, due to the different definitions of the convergence bispectrum, we get a factor of 3 difference with respect to S+05. We also use the symmetry of the bispectrum to only integrate from 0 to $\pi$ in $\varphi$, introducing a prefactor of 2.}
        \begin{align}
            \MapMapMap {}&{}(\theta_1,\theta_2,\theta_3) = \frac{2}{(2\pi)^3}\int_0^\infty \dd \ell_1\,\ell_1\int_0^\infty \dd \ell_2\,\ell_2 \int_0^\pi\dd\varphi \nonumber\\
            {}&{}\times\hat{u}(\theta_1\ell_1)\,
            \hat{u}(\theta_2\ell_2)\,\hat{u}\left(\theta_3\sqrt{\ell_1^2+\ell_2^2+2\ell_1\ell_2\cos\varphi}\right)\nonumber\\
            {}&{}\times b(\ell_1,\ell_2,\varphi)\; \label{eq:map3_from_bkappa}.
        \end{align}
        We use the public \textsc{cubature} library to solve this integration. In our implementation, the integration kernel is executed on a graphics processing unit (GPU), yielding a significant speed-up over parallelisation on central processing units (CPUs).
        
    \subsection{Measuring aperture mass statistics}
        \label{subsec:measuring_map}
        There are three main methods to estimate the third-order aperture mass statistics $\MapMapMap$, first, via the convergence field $\kappa$, second, via the shear field or, in practice, from the observed galaxy ellipticities, and third, via the third-order correlation functions $\Gamma^{(i)}$.
        
        \subsubsection{Measuring aperture mass statistics directly}
            \label{subsec:measuring_map_direct}
            The most straightforward way is to measure aperture mass maps directly on a convergence field using Eq.~\eqref{eq:definition_aperture_mass_kappa}. In a real survey, this is difficult, as the convergence is not directly observable. In principle, one could compute the aperture mass statistics of a reconstructed convergence field, but this is not a good way to estimate aperture statistics, as the convergence reconstruction yields a convergence map that is necessarily smoothed and potentially also inhibits other systematic effects. While not really applicable to real data, this method yields a quick and unbiased way to estimate aperture statistics in lightcones from simulations, as for them, convergence maps are readily available. However, one faces the issue of boundary effects when the integral in Eq.~\eqref{eq:definition_aperture_mass_kappa} extends past the simulation boundary. To avoid this issue, we cut off a slice of width $4\theta$ from the computed aperture mass maps\footnote{Both the $Q$- and the $u$-filter function have $99.9\%$ of their power within this range, meaning that boundary effects beyond this cut-off are negligible.}.

            Another way to estimate aperture statistics is from an ensemble of observed galaxy ellipticities, using 
            \begin{equation}
                \label{eq:map_discrete_estimator}
                {\MapEst}(\vec{\vartheta};\theta) + \ii {\MperpEst}(\vec{\vartheta};\theta) = \frac{1}{n_\mathrm{gal}}\sum_i Q_\theta(|\vec{\vartheta}-\vec{\vartheta}_i|)\left(\varepsilon_{\mathrm{t},i}+\ii\varepsilon_{\times,i}\right)\; ,
            \end{equation}
            where $\varepsilon_{\mathrm{t}/\times}$ are the observed galaxy ellipticities converted into their tangential and croos components according to Eq.~\eqref{eq:gamma_t_defn}; $\vec{\vartheta}_i$ are their respective positions. Here, $n_\mathrm{gal}$ can be the global number density of galaxies \citep{Bartelmann:2001} or the number density of galaxies within the aperture radius \citep{Martinet:2018}. For this work, we define $n_\mathrm{gal}$ as the number of galaxies weighted by the $Q$-filter function:
            \begin{equation}
                n_\mathrm{gal} = \sum_i Q_\theta(|\vec{\vartheta}-\vec{\vartheta}_i|) \; .
            \end{equation}
            We tested all three definitions of $n_\mathrm{gal}$ using the SLICS and found that, for randomly distributed galaxies, setting $n_\mathrm{gal}$ as the number density within the aperture radius or the one weighted by the $Q$-filter function induces sub-percent differences on the third-order aperture masses $\MapMapMap$. However, setting $n_\mathrm{gal}$ as the global galaxy density can induce differences of about $5\%$ in $\MapMapMap$.
            
            In the following, we adopt $n_\mathrm{gal}$ to be the number of galaxies weighted by the $Q$-filter function. Rewriting Eq.~\eqref{eq:gamma_t_defn} as
            \begin{equation}
                \gamma_t + \ii\gamma_\times = -\left(\gamma_1+\ii\gamma_2\right) \frac{(\vec{\vartheta}-\vec{\vartheta}')^*}{(\vec{\vartheta}-\vec{\vartheta}')}\; ,
            \end{equation}
            where the vector $\vec{\vartheta}-\vec{\vartheta}'$ denotes the projection direction of the tangential shear in complex notation, we can rewrite Eqs.~\eqref{eq:definition_aperture_mass_gamma} and \eqref{eq:map_discrete_estimator} as a convolution. To calculate aperture mass fields, we distribute galaxies on a grid using a cloud-in-cell method. From this, we compute the aperture masses using a Fast Fourier Transform (FFT), allowing us to compute an aperture mass map in $\mathcal{O}(\Npix\log\Npix)$ operations. To compute second- and third-order aperture statistics, we apply the same cut-off of $4\theta_\mathrm{ap}$ to the aperture mass maps. From these aperture mass maps, we obtain estimates for the second- and third-order aperture statistics by multiplication of the respective aperture mass maps on each pixel, and then taking the average of all pixel values.
            
            To extract the data vectors from the (full-sky) convergence maps of the T17 simulations, we smooth the maps with \textsc{healpy} function \textsc{smoothing}, with a given beam window function created by the function \textsc{beam$2$bl}, which in turn is determined by the corresponding $U_\theta$ filter. For each filter radius $\theta$, this yields a full-sky aperture mass map ${\MapEst}(\vec{\vartheta};\theta)$, without the need to cut off boundaries.

        \subsubsection{Measuring aperture statistics from three-point correlation functions}
            \label{subsec:measuring_map_from_3pcf}
            While the abovementioned method to estimate aperture statistics is extremely fast, it can not be applied to realistic survey data. Assuming a relatively large aperture radius of $\theta_\mathrm{ap}=30'$, we would have to cut off a $2\degree$-strip around every edge or mask in the survey footprint, meaning that we would disregard most of the data. While active research is being conducted to circumvent these problems \citep{Porth:2020,Porth:2021}, the arguably best method to estimate third-order aperture statistics from real data is to derive them from the measured 3pcf, as has been introduced in \citet{Jarvis:2004}, generalised in \citet{Schneider:2005} and applied to survey data in \citet{Fu:2014} and \citet{Secco:2022}. The shear 3pcf can be estimated straightforwardly from a survey with arbitrarily complex geometry, meaning that the converted aperture statistics are not biased by boundary effects. One caveat is that this conversion requires the knowledge of the 3pcf for all triangle configurations, particularly for infinitesimally small or extremely large ones, both of which can not be measured. The incomplete knowledge of the correlation functions can lead to a mixing of E- and B-modes for the aperture statistics \citep{Kilbinger:2006}. However, for third-order aperture statistics, this effect appears to be not as severe as for their second-order counterpart \citep[at least for the diagonal part of the aperture statistics, this has been demonstrated in][we are testing this assumption for non-diagonal aperture mass statistics in Sect.~\ref{subsec:validation_binning_choice}]{Shi:2014}. Despite its advantages, this method comes at the price of computation time: Calculating the shear 3pcf for a realistic number of source galaxies takes orders of magnitude longer\footnote{For a $10\times 10\,\mathrm{deg}^2$ field of a Stage-IV survey, direct estimation of aperture mass statistics takes a few minutes vs.~$\sim$1500 CPUh for the estimation of the 3pcf.} than the direct estimation of aperture statistics, so calculating the shear 3pcf of an ensemble of simulations (as would be necessary to estimate a covariance matrix) comes at a prohibitively high computational cost.
            
            The method to transform shear three-point correlation functions into aperture statistics is already implemented in \textsc{treecorr}. To compute the third-order aperture statistics, the quantities
            \begin{align}
                \expval{MMM}(\theta_1,\theta_2,\theta_3) = {}&{} A_1 \int \dd y_1 \int \dd y_2 \int_0^{2\pi}\dd \psi \nonumber\\
                \times {}&{} \Gamma^{(0)}_\mathrm{cen}(y_1,y_2,\psi) F_1(y_1,y_2,\psi)\;, \nonumber\\
                \expval{MMM^*}(\theta_1,\theta_2;\theta_3)= {}&{} A_2 \int \dd y_1 \int \dd y_2 \int_0^{2\pi}\dd \psi\label{eq:MMM_from_gamma}\\
                \times {}&{} \Gamma^{(3)}_\mathrm{cen}(y_1,y_2,\psi) F_2(y_1,y_2,\psi) \nonumber
            \end{align}
            need to be computed. Here, $A_{1,2}$ and $F_{1,2}$ are the prefactors and filter functions, which are specified in S+05 (see equations 62 and 71). The aperture statistics $\MapMapMap$ and $\MapMperpMperp$ are linear combinations of these quantities:
            \begin{align}
            	\MapMapMap {}&{}(\theta_1,\theta_2,\theta_3)
				 \nonumber\\=\,\,\,{}&{}\!\!\!\Re\left[\expval{M^2 M^*}(\theta_1,\theta_2;\theta_3)
				+\expval{M^2 M^*}(\theta_1,\theta_3;\theta_2) \right. \nonumber \\
				&\!\left.
				+\expval{M^2 M^*}(\theta_2,\theta_3;\theta_1)
				+\expval{M^3}(\theta_1,\theta_2,\theta_3)\right]/4 \;,\nonumber 
				\\
    			\MapMperpMperp {}&{}(\theta_1;\theta_2,\theta_3) \label{eq:map3_from_MMM}
				  \\=\,\,\,{}&{}\!\!\!\Re\left[\expval{M^2 M^*}(\theta_1,\theta_2;\theta_3)
				+\expval{M^2 M^*}(\theta_1,\theta_3;\theta_2) \right. \nonumber \\
				&\!\left.
				-\expval{M^2 M^*}(\theta_2,\theta_3;\theta_1)
				-\expval{M^3}(\theta_1,\theta_2,\theta_3)\right]/4 \;. \nonumber
            \end{align}
            We will not consider the quantities $\expval{\Map^2\Mperp}$ and $\expval{\Mperp^3}$, as they vanish for any parity-symmetric field \citep{Schneider:2003b}.
            In \textsc{treecorr}, this method is implemented in the following way: First the $\Gamma^{(1)}$ and $\Gamma^{(2)}$ are transformed into $\Gamma^{(3)}$ via Eq.~\eqref{eq:permutations_natural_components}. For each bin $(r,u,v)$, the transformation matrix $\frac{\dd\{r,u,v\}}{\dd\{y_1,y_2,\psi\}}$ is then calculated. The value of the respective integral is computed as the sum of the values of $\Gamma^{(3)}$ multiplied by the determinant of the transformation matrix and the value of the filter functions $F_{1,2}$ at the bin centre. While numerically very cheap, this is probably not the most efficient way to compute that integral. In case the filter functions vary significantly over a bin \citep[compare Fig.~2 and 3 of][]{Schneider:2005}, it might be more appropriate to calculate the average of the filter function in a bin, for example. We tried to improve the integration results by interpolating the measured shear 3pcf and performing the same integral, achieving rather moderate improvements. We leave an optimisation of the conversion from shear 3pcf to aperture mass statistics for future work.

    \subsection{Validation}
        \subsubsection{Binning choice of three-point correlation functions}
            \label{subsec:validation_binning_choice}
            \begin{figure*}
                \centering
                \includegraphics[width=17cm]{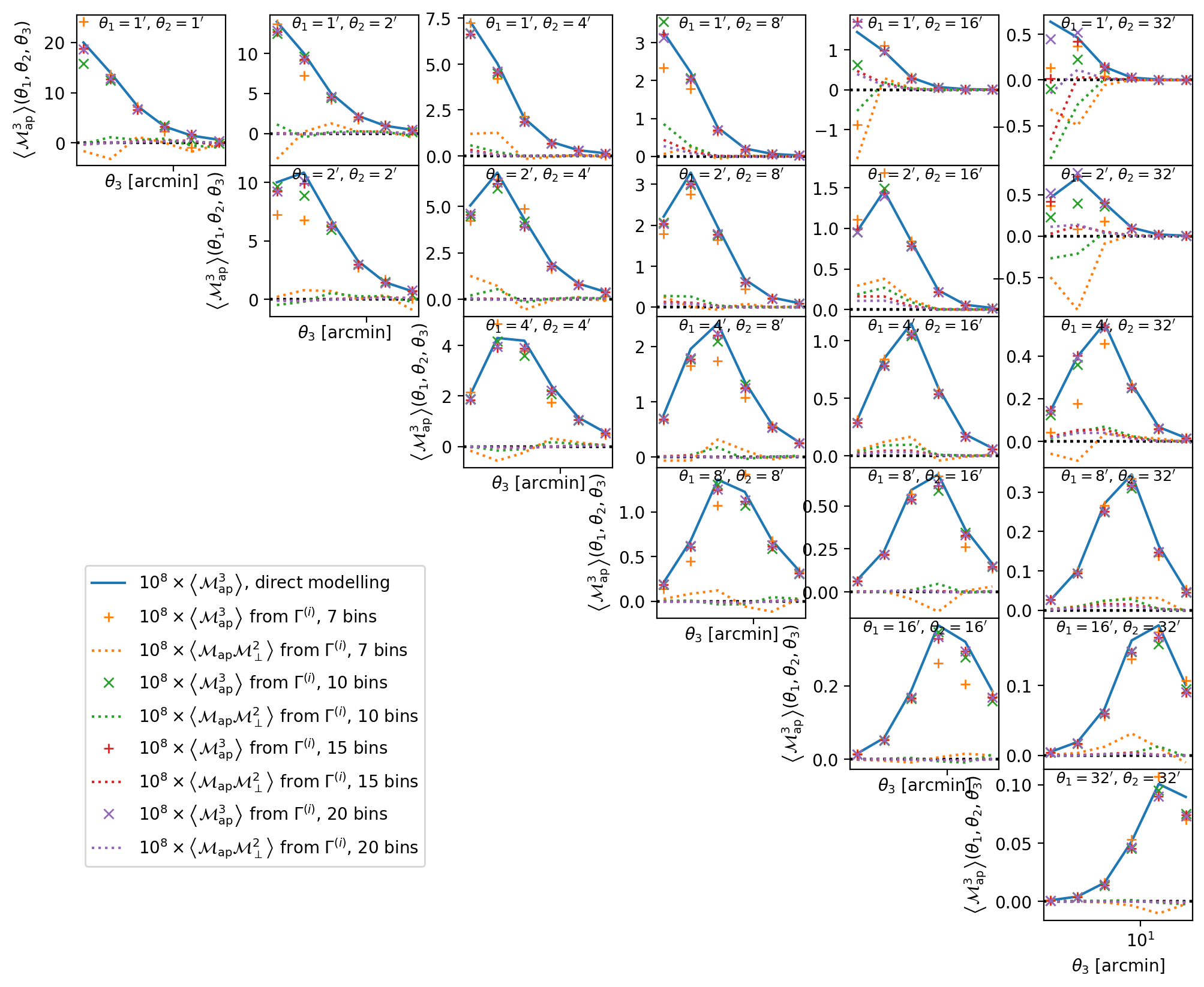}
                \caption{The third-order aperture mass statistics $\MapMapMap$ when modelled directly from the bispectrum (Eq.~\ref{eq:map3_from_bkappa}, blue line) compared to the ones constructed from modelled shear 3pcf (Eq.~\ref{eq:MMM_from_gamma}, coloured crosses). The dotted lines denote the respective B-modes. We plot $\MapMapMap(\theta_1,\theta_2,\theta_3)$, where $\theta_1$ is constant in each row, $\theta_2$ is constant in each column and $\theta_3$ varies along the $x$-axis in each panel.
                The yellow, green, red and purple dots denote the $\MapMapMap$ we get from the shear 3pcf when we bin $r$, $u$ and $v$ in $7\times 7\times 7$, $10\times 10\times 10$, $15\times 15\times 15$ and $20\times 20\times 20$ bins, respectively.}
                \label{fig:map3_from_gamma_binning}
            \end{figure*}
            
            As a first step, we want to validate the conversion $\Gamma^{(i)}\dashrightarrow\MapMapMap$ performed by the \textsc{TreeCorr} algorithm. In particular, we investigate the number of bins necessary to achieve an unbiased estimate for third-order aperture statistics and quantify the leakage of E- and B-modes. While the latter has already been investigated by \citet{Shi:2014}, we extend upon these results by using a realistic convergence bispectrum model and by taking into account non-diagonal aperture mass statistics.
            
            Our modelling pipeline provides us with the ability to test this conversion. As we start from the same convergence bispectrum $B_\kappa$, the aperture mass statistics achieved by the conversion $B_\kappa\overset{\eqref{eq:gamma0_from_bkappa}}{\to}\Gamma^{(i)}\overset{\eqref{eq:MMM_from_gamma}}{\dashrightarrow}\MapMapMap$ and by direct modelling $B_\kappa\overset{\eqref{eq:map3_from_bkappa}}{\to}\MapMapMap$ have to be consistent. Furthermore, the modelled $\Gamma^{(i)}$ are pure E-mode functions, so any B-modes $\MapMperpMperp$ that we observe have to be created by the transformation $\Gamma^{(i)}\overset{\eqref{eq:MMM_from_gamma}}{\dashrightarrow}\MapMapMap$. These tests would be unfeasible to perform with simulations due to the prohibitively high computational cost of extracting the shear 3pcf from an extensive simulation set for different bin sizes.
            
            The results of our tests can be seen in Fig.~\ref{fig:map3_from_gamma_binning}. We see that as long as the three filter radii $\theta_1$, $\theta_2$ and $\theta_3$ are similar, the conversion appears to work reasonably well. Only when we bin $\Gamma^{(i)}$ in $7^3$ bins do we get significant deviations, meaning that this is certainly not a sufficient number of bins. The conversion becomes less accurate when two filter radii are much smaller than the third one, as shown in the top-right corner of Fig.~\ref{fig:map3_from_gamma_binning}. There also the results for $10^3$ bins show significant deviations, whereas the results for $15^3$ bins seem overall consistent with the ones from $20^3$ bins. We also observe a non-negligible amount of B-mode leakage for these cases, even for $15^3$ and $20^3$ bins. We are planning to exclude all combinations of filter radii from cosmological parameter analyses where we observe a B-mode leakage of more than 10\% for the 3pcf in $15^3$ bins.
            
            \begin{figure}
                \centering
                \includegraphics[width=\linewidth]{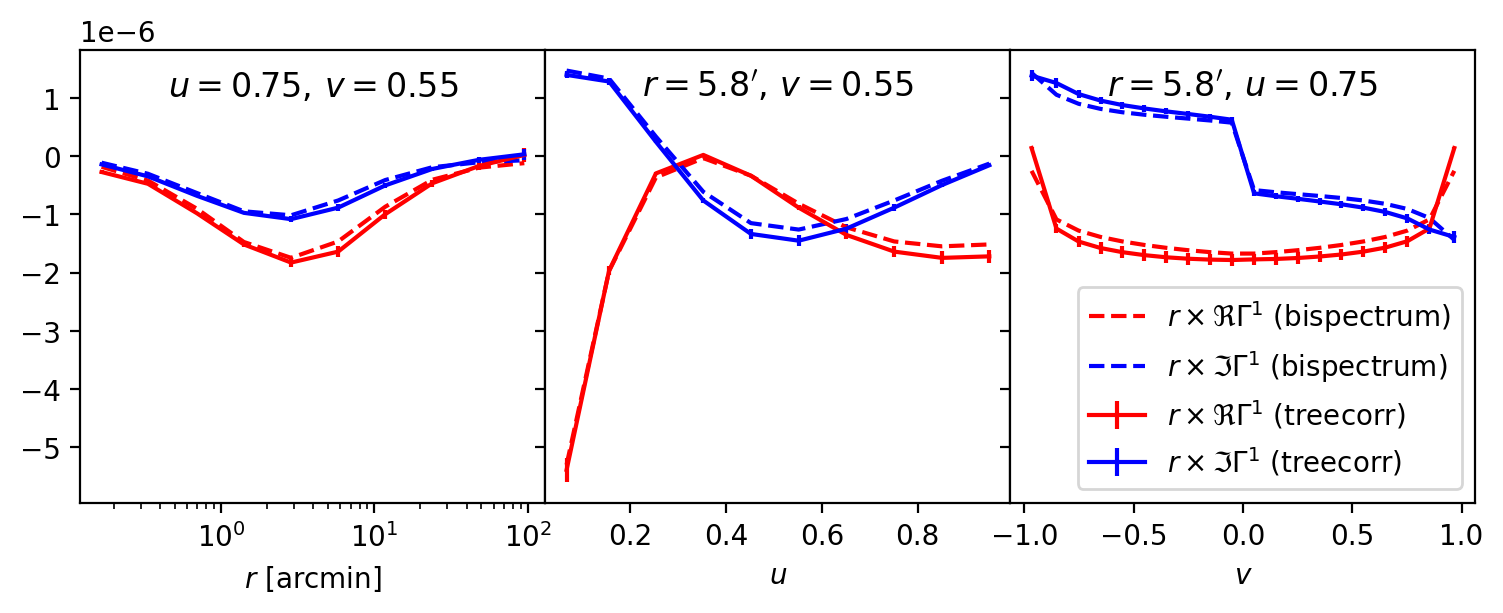}
                \caption{The three-point correlation function $\Gamma^{(1)}$ as a function of its three parameters $r$ (left), $u$ (middle) and $v$ (right). We note that the jump in the imaginary part of $\Gamma^{(1)}$ around $v=0$ is due to the fact that $\Gamma^{(i)}(r,u,v)=\Gamma^{(i)*}(r,u,-v)$ holds.}
                \label{fig:gamma1_of_r_u_v}
            \end{figure}
            Inspecting Fig.~\ref{fig:gamma1_of_r_u_v} we further note that the function $\Gamma^{(1)}(r,u,v)$ is relatively smooth and well-behaved with respect to $r$, but strongly varies as a function of $u$ and $v$, especially when $u\approx 0$ and $v\approx\pm 1$. This implies that $r$ can be binned rather coarsely, as long as $u$ and $v$ are finely binned. This is in contrast to binning choices in other studies, e.g.~\citet{Secco:2022}, who preferred a fine binning in $r$ (55 bins) and a coarser binning in $u$ and $v$ (10 bins).
        \subsubsection{Comparison to N-body simulations}
            \label{subsubsec:validation_n_body_sims_map}
            \begin{figure*}
                \centering
                \includegraphics[width=17cm]{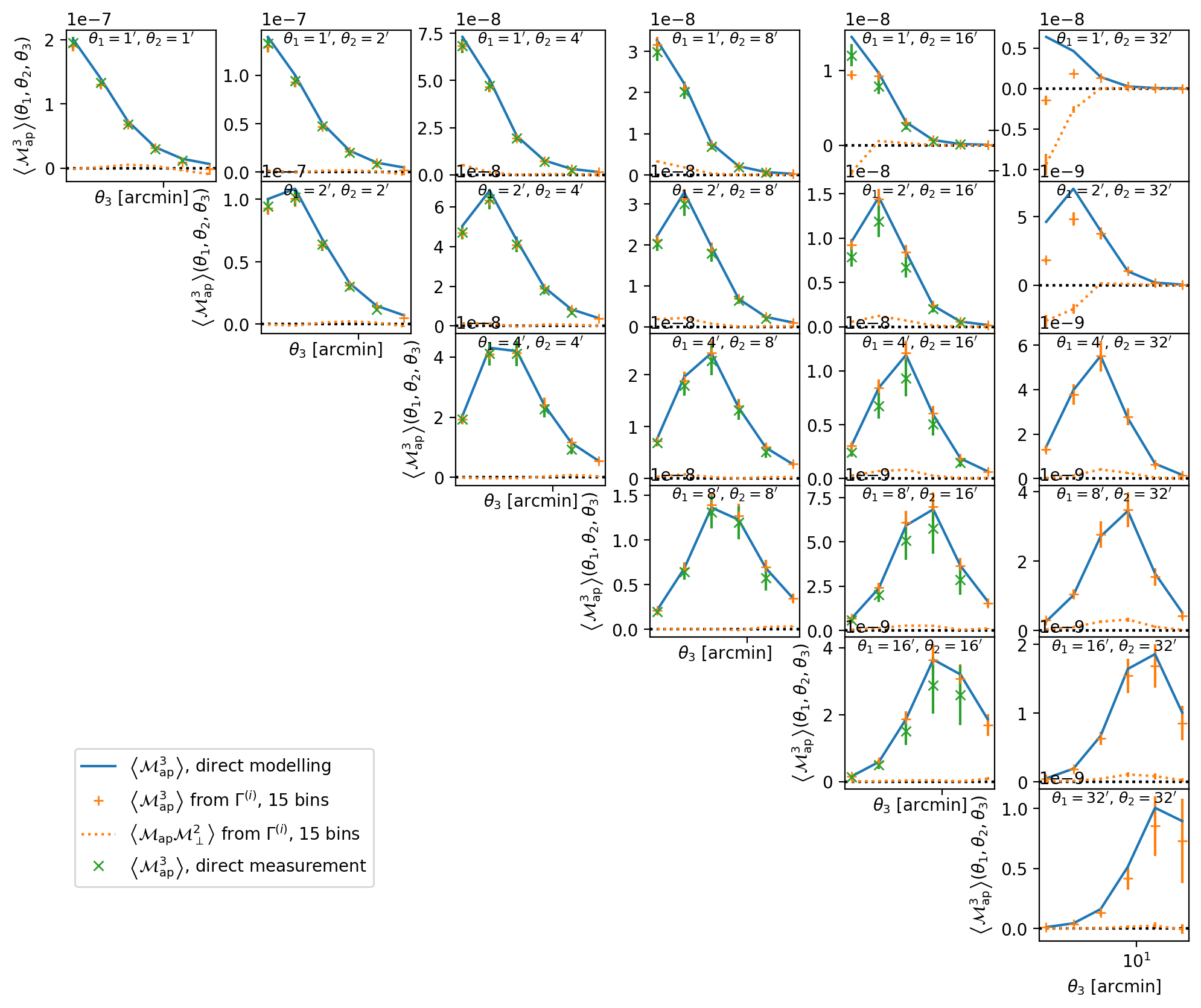}
                \caption{The third-order aperture masses from our modelling pipeline using \textsc{BiHalofit} (Eq.~\ref{eq:map3_from_bkappa},blue) compared to the ones measured directly from the MS. The direct measurements on the convergence maps (Eq.~\ref{eq:definition_aperture_mass_kappa}) are shown in green, and the measurements converted from shear 3pcf (Eq.~\ref{eq:MMM_from_gamma}) are shown in orange. Both statistics were computed from 32 lines of sight of the MS, the error bars denote the error on the mean. We note that for the largest angle of $\astroang{;32;}$, a direct measurement could not be obtained due to the limited size of the individual light cones.}
                \label{fig:third_order_map_model_vs_MS}
            \end{figure*}
            We compare the modelled aperture mass statistics with the ones we measure in the MS in Fig.~\ref{fig:third_order_map_model_vs_MS}. As expected from our discussions in Sect.~\ref{subsec:validation_binning_choice}, we observe that the conversion from shear 3pcf to third-order aperture masses fails when two aperture radii are small, and the third one is large (as observed in the top-right panel). In these cases, we also register significant B-modes. We also note that in most cases, the uncertainties on the direct measurements are significantly larger than those from the shear 3pcf. Cutting off a stripe around the boundary in order to avoid edge effect in the estimate of $\MapMapMap$ (as discussed in Sect.~\ref{subsec:measuring_map}), leads to a loss of information compared to measuring the 3pcfs, for which all triplets of available points\footnote{Again, we only use every tenth pixel to calculate the 3pcf, but do not expect the a strong loss of signal to noise from this.} in the field are used.

\section{Cosmological parameter estimation}
\label{sec:results_mcmc}
To perform a cosmological parameter inference, we use our described pipeline to create a model vector. For the covariance, we rely on N-body simulations, where we use the method of \cite{Percival2021} to debias the estimated covariance matrix $\Tilde{C}$. Given a data vector $\vb{d}$ and a covariance matrix $\Tilde{C}$ measured from $n_\mathrm{r}$ simulated survey realisations, the posterior distribution of a model vector $\vb{m}(\boldsymbol{\Theta})$ depending on $n_\Theta$ parameters, is
\begin{equation}
    \boldsymbol{P}\left(\vb{m}(\boldsymbol{\Theta})|\vb{d},\Tilde{C}\right) \propto |\Tilde{C}|^{-\frac{1}{2}} \left( 1 + \frac{\chi^2}{n_{\rm r}-1}\right)^{-m/2}\, ,
    \label{eq:t_distribution}
\end{equation}
where
\begin{equation}
\chi^2 =  \left[\vb{m}(\boldsymbol{\Theta})-\vb{d}\right]^{\rm T} \Tilde{C}^{-1} \left[\vb{m}(\boldsymbol{\Theta})-\vb{d}\right] \, .
\label{eq:chi2}
\end{equation}
The power-law index $m$ is 
\begin{equation}
    m = n_\Theta+2+\frac{n_\mathrm{r}-1+B(n_\mathrm{d}-n_\Theta)}{1+B(n_\mathrm{d}-n_\Theta)} \;,
    \label{eq:m_exponent}
\end{equation}
with $n_{\rm d}$ being the number of data points and
\begin{equation}
    B = \frac{n_\mathrm{r}-n_\mathrm{d}-2}{(n_\mathrm{r}-n_\mathrm{d}-1)(n_\mathrm{r}-n_\mathrm{d}-4)} \, .
    \label{eq:B}
\end{equation}
If $m=n_\mathrm{r}$ the formalism of \cite{Sellentin2016} is recovered.

Normally, one needs to evaluate this likelihood function in high-dimensional parameter space to perform a cosmological parameter analysis with third-order statistics. 
The creation of the model vector for third-order aperture statistics with 35 combinations of aperture radii takes about one minute on an NVIDIA A40 GPU, and needs to be evaluated at about $10^4$ points even for sampling methods like \textsc{polychord} \citep{Handley:2015}. This means that a complete likelihood analysis with third-order aperture statistics is possible but takes a long time, whereas a complete likelihood analysis with the 3pcf, where the modelling of the 3pcf in $10^3$ bins takes two to three hours, is not feasible. To solve this issue, we use a neural network emulator called \textsc{CosmoPower} \citep{COSMOPOWER2022}, which was first developed to emulate power spectra but can easily be adapted for arbitrary vectors. Since a neural network emulator needs as many as possible evaluation points, we calculate our model at 7500 points\footnote{Due to the long modelling time, we only use 500 of the 7500 sampled points for the 3pcf.} in a four-dimensional Latin hypercube describing a flat $w$CDM cosmological model, varying the parameters $\Omm, S_8, w_0$ and $h$. We leave all other parameters fixed at the values corresponding to the SLICS, which we introduced in Sect.~\ref{sec:nbody_sims}. We then train the emulator with 6500 points and use the remaining 1000 as a validation test, as shown in Fig.\ref{fig:emulator_acc}. This neural network-based emulator performs extraordinarily well, modelling all statistics with a sub-per-cent accuracy. As our bispectrum model is only accurate to about 10\% \citep{Takahashi:2020}, the emulator uncertainty plays a negligible role in the modelling process.

\subsection{Shear three-point correlation functions vs.~third-order aperture masses}
One aspect of this work is investigating the information loss using aperture mass statistics instead of the 3pcf itself. Although the aperture mass statistics are well suited for a cosmological parameter inference due to their E-/B-mode decomposition and fast modelling times, they should not be used if the loss of information is too severe.
        
Unfortunately, we cannot quantify the full information content of the shear 3pcf, as the data vector contains about $10^4$ entries, and therefore a reliable covariance matrix is not accessible. To circumvent this problem, we model the shear 3pcf in $10^3$ bins in $r,u$ and $v$, where we bin $r$ logarithmically from $0.\!'1$ to $\astroang{;100;}$, at 500 of the 7500 training nodes. We then perform a principal component analysis (PCA) to decide on the 40 most relevant principal components of the 3pcf data vector.

Using this PCA, we determine the covariance matrix for the shear 3pcf, which we measure from 200 $10\times 10\,\mathrm{deg}^2$ tiles of the SLICS in the same configuration as the training data. For the model vector, we again use \textsc{CosmoPower}, which is trained on the PCA components of the 500 models used in determining the principal components.

\begin{figure}
\begin{subfigure}{\columnwidth}
\includegraphics[width=\columnwidth]{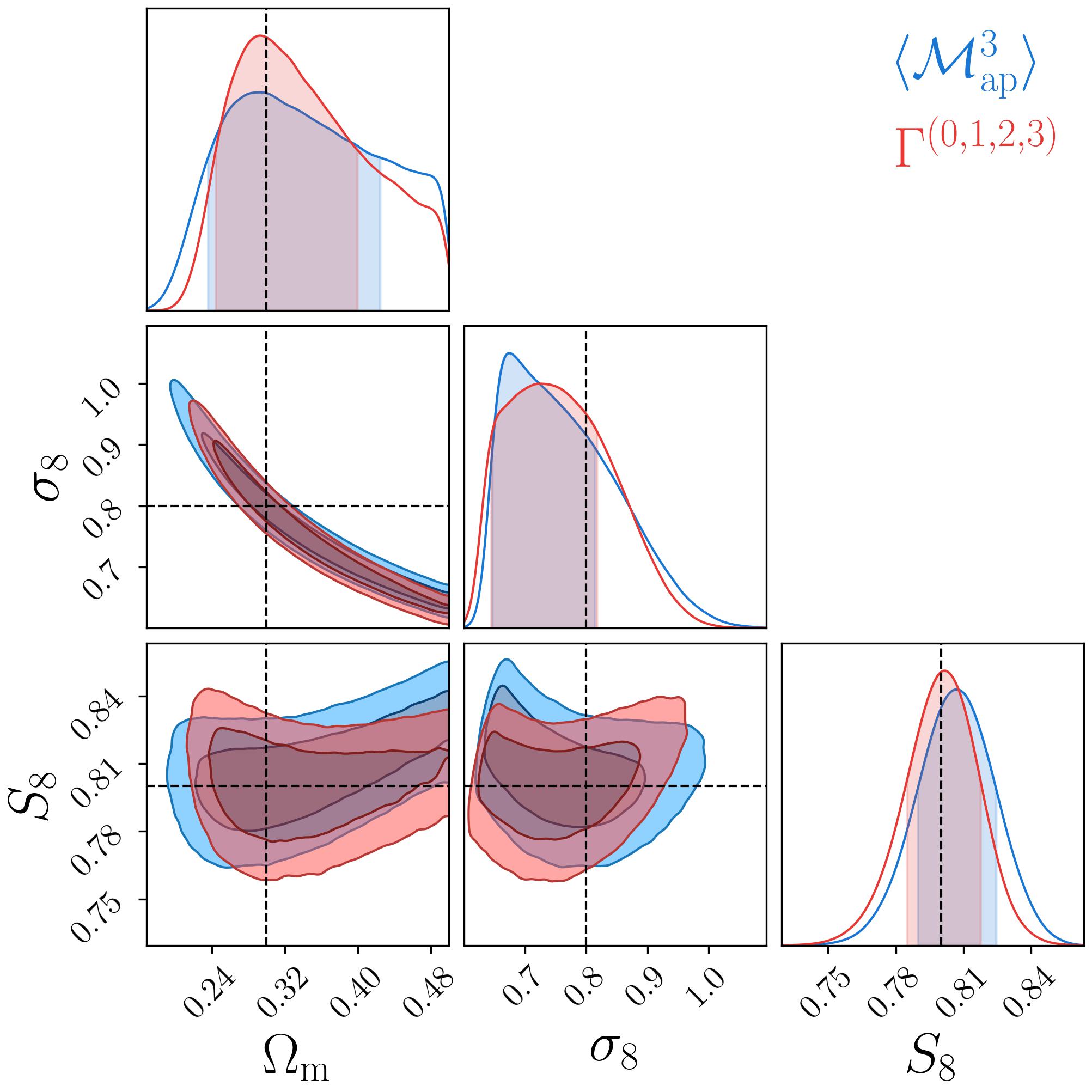}
\end{subfigure}
\caption{Comparison of the posterior distributions of different third-order statistics using the SLICS simulations to determine the covariance. In red we show the posteriors of the PCA from the 3pcf and in blue the one resulting from the $\langle \mathcal{M}_\mathrm{ap}^{3} \rangle$ statistic.}
\label{fig:MCMC_Map3vsGamma}
\end{figure}
    
In Fig.~\ref{fig:MCMC_Map3vsGamma} we compare the constraining power of the PCA analysis of 3pcf to the $\langle \mathcal{M}_\mathrm{ap}^{3} \rangle$ analysis, where the covariance for PCA analysis of 3pcf is measured from 200 SLICS realisations; for $\langle \mathcal{M}_\mathrm{ap}^{3} \rangle$ we use all $927$ available SLICS realisations and take into consideration all combinations of the filter radii $\astroang{;0.5;},\astroang{;1;},\astroang{;2;},\astroang{;4;},\astroang{;8;},\astroang{;16;}$, and $\astroang{;32;}$. The different number of realisations are considered by Eq.~\eqref{eq:m_exponent}. In both cases, the data vector was created by the \textsc{CosmoPower} Emulator. It is clearly seen that the constraining power from the PCA analysis of 3pcf is only slightly better than the one from $\langle \mathcal{M}_\mathrm{ap}^{3} \rangle$, and this slight difference may very well be explained by the use of different scales between the 3pcf and aperture mass statistics, and the fact that the aperture mass maps are smaller due to the cutoff at the boundaries. Overall, the advantages of the $\langle \mathcal{M}_\mathrm{ap}^{3} \rangle$ justify their use, even considering their potentially slightly lower constraining power.   
        
\subsection{Combination of second- and third-order aperture masses}
To assess the constraining power for third-order aperture statistics, especially when combined with second-order shear statistics, we perform a mock analysis for a non-tomographic KiDS-1000-like setup.
            
To estimate the covariance matrix, we made use of all 108 realisations of the T17 simulations with a resolution $\textsc{nside}=4096$ (corresponding to a pixel size of $\astroang{;0.74;}^2$). From each realisation, we extract 18 HEALPix squares of size $\approx 860\,\mathrm{deg}^2$ that do not share common borders. This results in 1944 independent realisations from which the covariance matrix is estimated, such that with a data vector size of $\sim 50$, the statistical noise of the covariance matrix can be neglected. Since the area of the square patches is slightly larger than the one from KiDS-1000 with $\approx 777.4\,\mathrm{deg}^2$, the covariance needs to be re-scaled by a factor of $1.11$.
    
Furthermore, in order to have a data vector as unbiased and noise-free as possible, we estimate it with one full-sky realisation with a resolution $\textsc{nside}=8192$ (corresponding to a pixel size of $0.18$\,arcmin). Lastly, we use the filter scales of $(4,8,16,32)$\,arcmin, as the model and simulations are inconsistent for smaller scales.

\begin{figure*}
\begin{subfigure}{\columnwidth}
\includegraphics[width=\columnwidth]{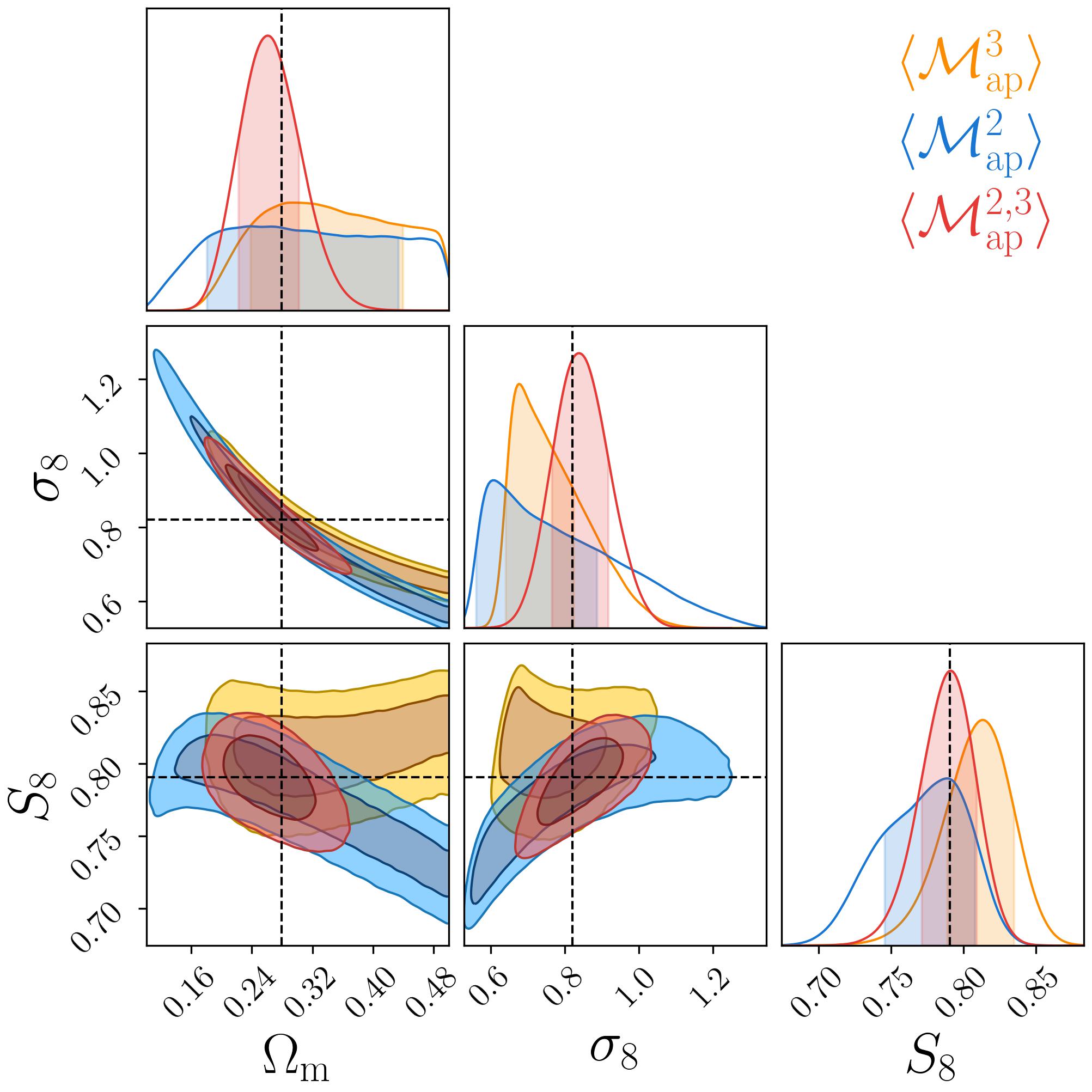}
\end{subfigure}
\begin{subfigure}{\columnwidth}
\includegraphics[width=\columnwidth]{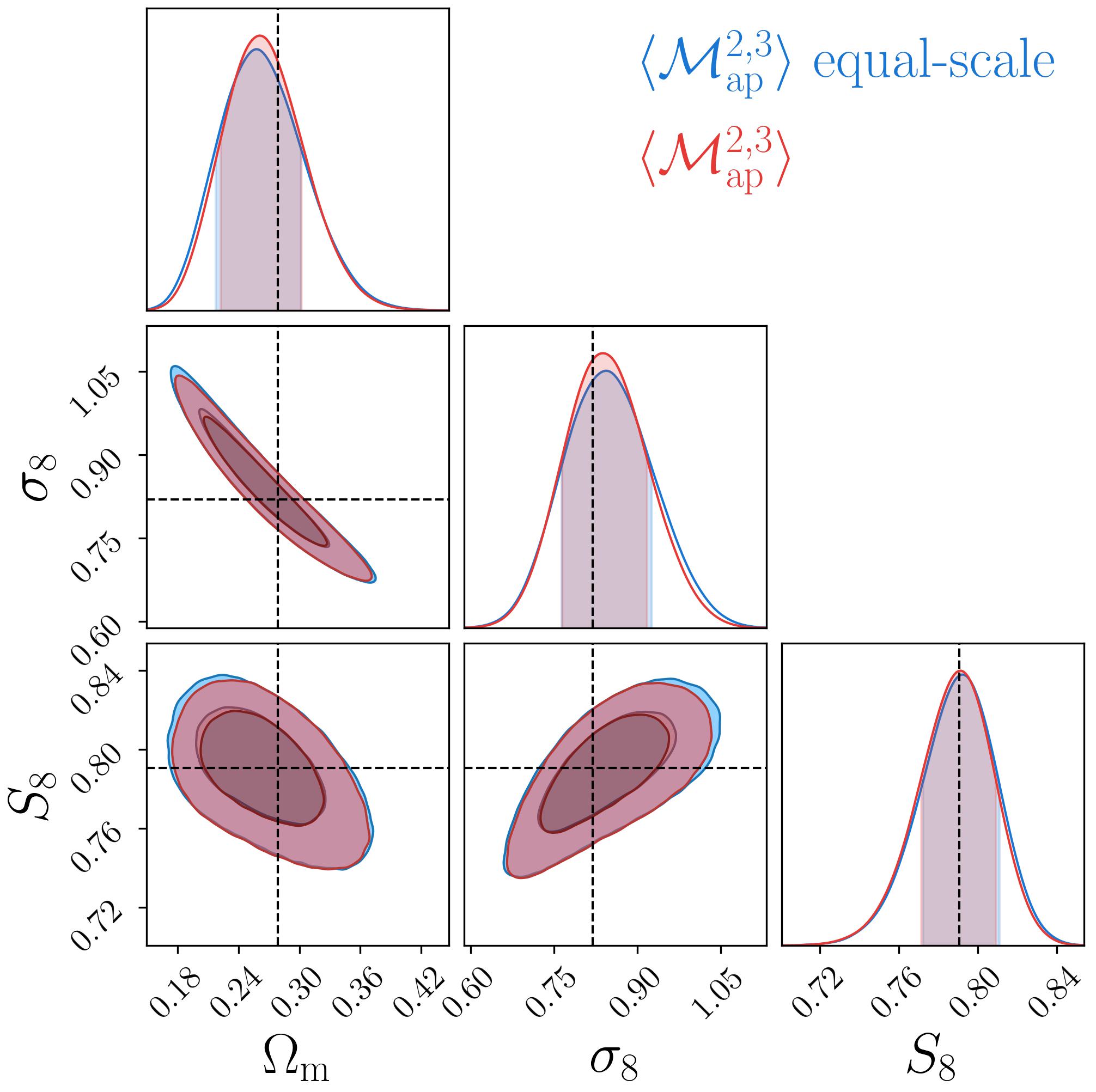}
\end{subfigure}
\caption{The left figure compares the posterior distributions of second- and third-order aperture statistics with a joint analysis within a KiDS-1000-like setup. The data vector and covariance matrix are estimated by the T17 simulations, and we use filter scales of $(4,8,16,32)$\,arcmin. The Hubble parameter $h$ and the dark energy equation-of-state are fixed to the T17 values. The right figure compares the combined second- and third-order aperture statistics with the combination of second-order aperture statistics with the equal-scale third-order aperture statistics. The corresponding data vector and model vector can be found in Fig.~\ref{fig:Map23_vector}. }
\label{fig:MCMC_T17}
\end{figure*}

The resulting posterior distribution is shown in the left panel of Fig.~\ref{fig:MCMC_T17}, where we first notice that the combination of second- and third-order statistics significantly increases the constraining power, especially in the $\Omega_\mathrm{m}$-$\sigma_8$ panel to the different degeneracy directions of the individual statistics (compare Tab.~\ref{tab:parameter_constraints}). Indeed, a joint analysis increases the constraints on $S_8$ by 42\% with respect to second-order statistics; the constraints on $\Omm$ and $\sigma_8$ increase by at least 68\% and 54\%, respectively\footnote{As the constraints for second-order statistics are at least partly dominated by the prior, the true increase is likely to be much greater}. The figure-of-merit \citep{Albrecht:2006} in the $\Omm$-$\sigma_8$ plane increases by a factor of 5.88. Finally, we note that the true cosmology is well within $1\,\sigma$ of the expected KiDS-1000 uncertainty for all three statistics.

Additionally, we investigate the constraining power if only equal-scale aperture masses $\MapMapMap(\theta,\theta,\theta)$ are used. As displayed in the right panel of Fig.~\ref{fig:MCMC_T17}, the loss of constraining power by the limitation to equal-scale aperture masses is small, although not zero. This is in stark contrast to \citet{Kilbinger:2005}, who found a strong difference in constraining power using a Fisher forecast. However, their analysis was conducted using a covariance matrix from a significantly smaller set of simulations. Our results are roughly in line with the findings of \citet{Fu:2014}, who found rather marginal differences in an MCMC. We note that these findings might change when additional parameters (either cosmological or nuisance) are introduced in the MCMC. In that case the equal-scale aperture masses will suffer from some degeneracies that the aperture mass statistics containing all filter radii might be able to break.

\begin{table}[]
    \centering
    \caption{Marginalised one-dimensional parameter constraints from the left part of Fig.~\ref{fig:MCMC_T17}. For $\expval{\Map^2}$ and $\expval{\Map^3}$ we do not cite upper limits on $\Omm$ or lower limits on $\sigma_8$ as they are dominated by the prior}
    \begin{tabular}{c|c c c}
        parameter & $\expval{\Map^2}$ & $\expval{\Map^3}$ & $\expval{\Map^{2,3}}$ \\
        \hline
         & & & \\[-0.8em]
$\Omm$ & $0.294_{-0.059}$ & $0.229_{-0.048}$ & $0.26^{+0.041}_{-0.04}$ \\[0.1em]
$\sigma_8$ & $0.671^{+0.155}$ & $0.603^{+0.287}$ & $0.842^{+0.075}_{-0.074}$ \\[0.1em]
$S_8$ & $0.813^{+0.023}_{-0.024}$ & $0.786^{+0.022}_{-0.041}$ & $0.792^{+0.017}_{-0.019}$
    \end{tabular}
    \label{tab:parameter_constraints}
\end{table}

\section{Discussion}
    \label{sec:discussion}
    In this work, which is the first of a series on cosmological analysis with third-order shear statistics, we have shown that a cosmological parameter analysis with third-order aperture mass statistics is feasible and beneficial for Stage-III surveys.
    
    Both the shear 3pcf and the third-order aperture statistics can be modelled from the matter bispectrum, and we found that our models based on the \textsc{BiHalofit} bispectrum model are accurate enough for Stage-III surveys. In particular, the flat-sky and Limber approximations are valid for our selected range of scales, so the accuracy of our model is mainly limited by the accuracy of the bispectrum model. We note that we have not yet tested the impact of astrophysical or observational systematics.

    We developed a test for binning strategies of the three-point correlation functions in order to obtain unbiased estimates for aperture mass statistics and found that, at our selected scales, a measurement of the three-point correlation functions in $15^3$ bins yields good results. In particular, we found that the leakage between E- and B-modes is at the percent level, and the bias in the aperture mass statistics is well below the sample variance for a Stage-III survey. This extends upon the findings of \citet{Shi:2014}, who found a percent-level leakage for diagonal aperture mass statistics utilising a simplified bispectrum model. We emphasise that, in addition to the effect of a minimum scale in the shear 3pcf investigated by \citet{Shi:2014}, our approach also quantifies the leakage that stems from the binning choices in the shear 3pcf, resulting in an inaccurate evaluation of the integral in Eq.~\eqref{eq:MMM_from_gamma}.
    
    We have tested the information loss when converting the shear three-point correlation functions to third-order aperture mass statistics by performing a principal component analysis of the three-point correlation functions and comparing the constraining power of the principal components to the one of third-order aperture statistics. We found comparable information content, suggesting that third-order aperture statistics constitute a good data compression method for the shear three-point correlation functions. In addition to an easier modelling, the aperture statistics have the added advantage that they cleanly separate E- and B-modes.
    
    We demonstrate that the computational load of a cosmological parameter analysis with third-order aperture statistics is manageable, particularly when utilising an emulator to speed up the MCMC. We make a \textsc{cosmoSIS}-module of our modelling algorithm publicly available at \url{https://github.com/sheydenreich/threepoint/releases}.
    
    Finally, we compare the constraining power between second-order, third-order, and joint aperture statistics analysis. While second-order aperture statistics are not being used in modern cosmological parameter analyses, we assume that all second-order shear statistics exhibit similar constraining power and parameter degeneracies \citep[compare][]{Asgari:2021}. We find that third-order aperture statistics alone have a lower constraining power than their second-order counterpart, but they exhibit a different degeneracy direction in the $\Omm$--$\sigma_8$-plane so that a joint analysis almost doubles the constraining power on the structure growth parameter $S_8$ and increases the figure-of-merit in the $\Omm$--$\sigma_8$-plane by a factor of 5.9. However, the information gain predicted by Fisher analyses, especially for the difference between diagonal and full third-order aperture statistics \citep[compare][]{Kilbinger:2005}, appears overly optimistic. This suggests that a Fisher forecast might not be an optimal tool to forecast parameter constraints and mock sampling methods give more realistic constraints.
    
    While we have demonstrated here that cosmological analyses with third-order shear statistics are feasible and promising, there are steps left to do before applying our methods to a concrete cosmological survey. The first of these is the development of a model for the covariance of third-order statistics, which is essential for a tomographic analysis and will be addressed in the following paper of this series (Linke et al., in prep). Additionally, as mentioned above, our model does not yet incorporate systematic and astrophysical effects, like baryonic feedback or intrinsic alignments of source galaxies \citep{Semboloni:2013,Pyne:2022}. We will develop and test strategies for treating these effects in future works of this series.

\begin{acknowledgements}
We thank Benjamin Joachimi and Mike Jarvis for providing valuable insights to this project. We would like to thank Joachim Harnois-D\'eraps for making public the SLICS mock data, which can be found at \url{http://slics.roe.ac.uk/}.
This work was funded by the TRA Matter (University of Bonn) as part of the Excellence Strategy of the federal and state governments. This work has been supported by the Deutsche Forschungsgemeinschaft through the project SCHN 342/15-1. SH acknowledges support from the German Research Foundation (DFG SCHN 342/13), the International Max-Planck Research School (IMPRS) and the German Academic Scholarship Foundation.\\
\emph{Author contributions.} All authors contributed to the development and writing of this paper. SH wrote the pipeline to model the bispectrum and shear 3pcf and the methods to measure the third-order statistics. LL implemented the modelling algorithm for aperture mass statistics, the GPU-integration, and the cosmosis module, aside from making various improvements to the codes. PB was responsible for everything regarding the T17 simulations and the MCMC runs, including the \textsc{CosmoPower} emulator. PS gave countless valuable insights into third-order shear statistics.
\end{acknowledgements}

\bibliographystyle{aa}
\bibliography{cite}

\appendix
\onecolumn
\section{Testing the \textsc{BiHalofit} bispectrum model}
    \label{sec:app_testing_bispectrum}
    \subsection{Measuring the bispectrum}
        \label{subsec:measuring_bispectrum}
        To measure the convergence bispectrum $\bkappa$ from simulations, we adapt the estimator developed by \citet{Watkinson:2017}. While their algorithm has been presented for three-dimensional density fields, it can be adapted to two-dimensional convergence fields.

        Given a convergence field $\kappa(\vtheta)$ and its Fourier transform $\hat{\kappa}(\vell)$, for an $\ell$-bin $\bar{\ell}_i = [\ell_\mathrm{min},\ell_\mathrm{max}]$ we define $\hat{\kappa}(\vell;\bar{\ell}_i)$ as 
        \begin{equation}
            \hat{\kappa}(\vell;\bar{\ell}_i) = \begin{cases}
            \hat{\kappa}(\vell) \qquad & \ell_\mathrm{min}\leq |\vell| < \ell_\mathrm{max} \\
            0 & \mathrm{otherwise}
            \end{cases}\; ,
            \label{eq:defn_kappa_cut}
        \end{equation}
        and $\kappa(\vtheta;\bar{\ell}_i)$ as its inverse Fourier transform. We also define $I(\vtheta;\bar{\ell}_i)$ as the inverse Fourier transform of $\hat{I}(\vell;\bar{\ell}_i)$ with $\hat{I}$ defined as in Eq.~\eqref{eq:defn_kappa_cut}:
        \begin{equation}
            \hat{I}(\vell;\bar{\ell}_i) = \begin{cases}
            1 \qquad & \ell_\mathrm{min}\leq |\vell| < \ell_\mathrm{max} \\
            0 & \mathrm{otherwise}
            \end{cases}\; .
            \label{eq:defn_kappa_cut_2}
        \end{equation}
        The estimator for the convergence bispectrum is then defined as 
        \begin{equation}
            \bkappa(\bar{\ell}_1,\bar{\ell}_2,\bar{\ell}_3) = \frac{\Omega^2}{\Npix^3}\frac{\sum_i^{\Npix} \kappa(\vtheta_i;\bar{\ell}_1)\kappa(\vtheta_i;\bar{\ell}_2)\kappa(\vtheta_i;\bar{\ell}_3)}{\sum_i^{\Npix} I(\vtheta_i;\bar{\ell}_1)I(\vtheta_i;\bar{\ell}_2)I(\vtheta_i;\bar{\ell}_3)} \; ,
        \end{equation}
        where $\Omega$ is the solid angle of the respective field and $\Npix$ is the number of pixels. The advantage of this estimator is its speed: With seven Fourier transforms, we can extract the complete averaged bispectrum of a field, where the three Fourier transforms required for the computation of $I(\vtheta;\bar{\ell_i})$ only need to be performed once, even when computing the bispectra of multiple fields. Furthermore, $\kappa(\vtheta;\bar{\ell}_i)$ can be stored for computing bispectra of different triangle configurations containing the same $\bar{\ell}$-bin. Nevertheless, the estimator suffers from one significant drawback: The Fourier transform assumes periodicity of the field $\kappa$, which is normally not given for convergence maps (in contrast to the three-dimensional $N$-body simulation cubes, which usually exhibit periodic boundary conditions). Therefore the estimator can be biased for $\ell$-scales approaching the scales of either the field or individual pixels. We, therefore, discard scales smaller than 5 pixels or larger than a third of the field size.
        
    \subsection{Validation}
        \label{subsubsec:validation_n_body_sims_bispectrum}
        \begin{figure*}
            \centering
            \sidecaption
            \includegraphics[width=12cm]{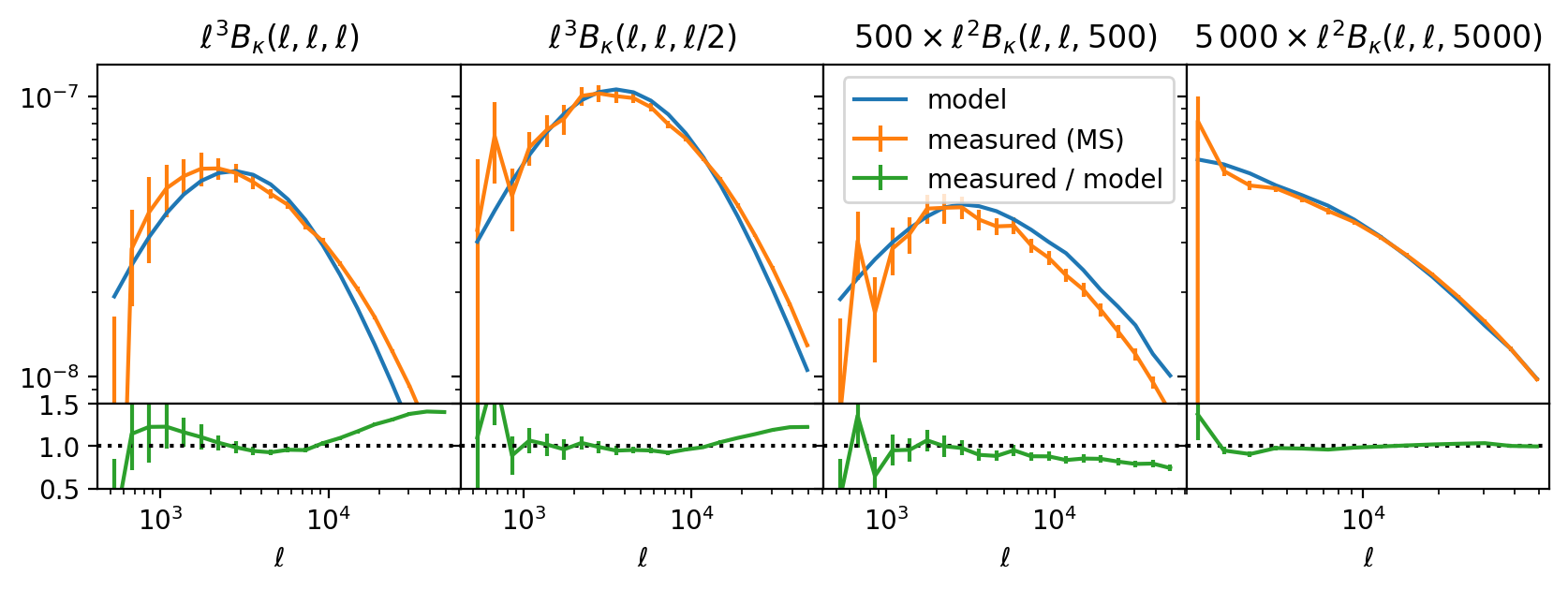}
            \caption{In the top panels we show the convergence bispectrum at $z=1$ extracted from the MS (orange) and modelled from \textsc{BiHalofit} (blue). The bottom panels show the ratio of measurement over the model. The error bars denote the error on the mean of all 64 lines of sight. For each point, we compute the average in a bin of width $\Delta\ell=0.13\ell$, both in the measurements and the model. For the model, we use equation (8) of \citet{Joachimi:2009} as a weight for different triangle configurations within one bin.}
            \label{fig:bispectrum_MS_model_vs_theory}
        \end{figure*}
        To validate our implementation of the \textsc{BiHalofit} algorithm and the Limber integration, we compare our model bispectrum with one extracted from the MS. For all 64 lines-of-sight, we take a convergence map at redshift $z=1$ and use the estimator described in Sect.~\ref{subsec:measuring_bispectrum} to extract the bispectrum for a set of triangle configurations. We compare these to our model predictions in Fig.~\ref{fig:bispectrum_MS_model_vs_theory}. We recover the bispectrum signal quite well, although we fall short of the 10-20\% accuracy reported in \citet{Takahashi:2020}. However, that may very well be due to the limited sample size provided in the MS or the smoothing inside the MS induced by the ray-tracing \citep{Hilbert:2009}.
\section{Testing the integration routine for $\Gamma_i$}
\label{sec:app_test_gamma_integration}
\begin{figure*}
      \centering
      \includegraphics[width=17cm]{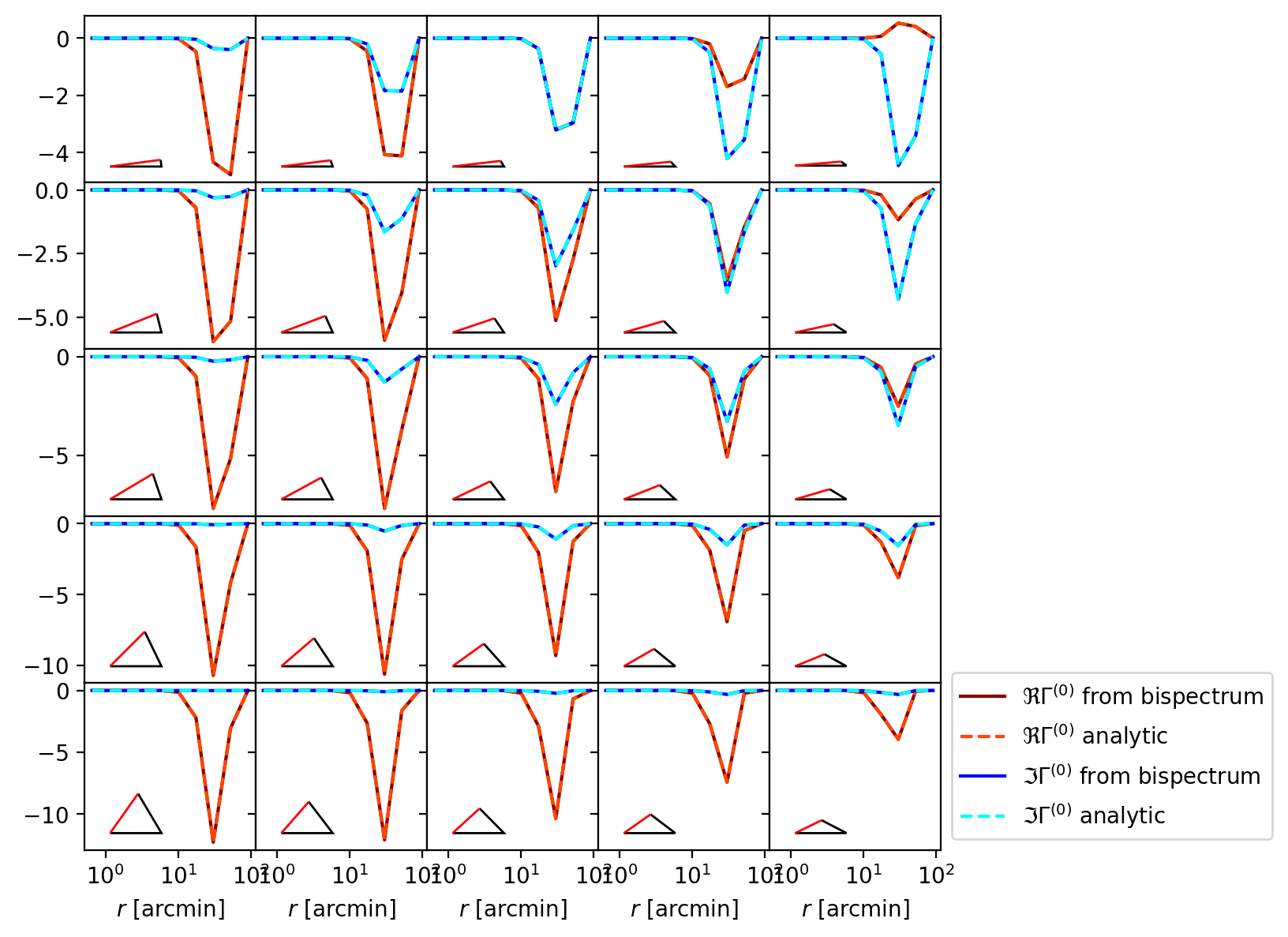}
      \caption{The real (red) and imaginary (blue) parts of the first natural component of the shear three-point correlation function $\Gamma^{(0)}$. Each plot panel corresponds to one fixed triangle shape shown in the bottom-right corner; the $x$-axis represents the length of the red triangle shape. We compare an integration of the analytic bispectrum model (\ref{eq:bkappa_analytic_model}, dark solid) and the analytic solution (light dashed). \eqref{eq:gammagammagamma_analytic_model}.}
      \label{fig:shear_3pcf_analytic_model}
\end{figure*}
When we define the three-point correlation function of the deflection potential as
\begin{equation}
      \myexpval{\psi(\vec{X})\psi(\vec{Y})\psi(\vec{Z})} = \frac{1}{8\alpha^3}\ee^{-\alpha\left[(\vec{X}-\vec{Y})^2+(\vec{Y}-\vec{Z})^2+(\vec{X}-\vec{Y})^2\right]}\; ,
\end{equation}
we can analytically compute both the three-point correlation functions and the bispectrum. For this, we define
\begin{equation}
    \partial_X = \partial_{X_1} + \ii \partial_{X_2} \; , \quad \nabla^2_X = \partial_X\partial^*_X \; ,
\end{equation}
and use the relations
\begin{align}
    \myexpval{ \kappa(\vec{X})\kappa(\vec{Y})\kappa(\vec{Z}) }= & \left(\frac{1}{2}\nabla^2_X\right)\left(\frac{1}{2}\nabla^2_Y\right)\left(\frac{1}{2}\nabla^2_Z\right) \myexpval{ \psi(\vec{X})\psi(\vec{Y})\psi(\vec{Z})} \; ,\\
    \myexpval{ \gamma(\vec{X})\gamma(\vec{Y})\gamma(\vec{Z})} = & \left(\frac{1}{2}\partial^2_X\right)\left(\frac{1}{2}\partial^2_Y\right)\left(\frac{1}{2}\partial^2_Z\right) \myexpval{ \psi(\vec{X})\psi(\vec{Y})\psi(\vec{Z})} \; ,\\
    \myexpval{ \gamma(\vec{X})\gamma(\vec{Y})\gamma^*(\vec{Z})} = & \left(\frac{1}{2}\partial^2_X\right)\left(\frac{1}{2}\partial^2_Y\right)\left(\frac{1}{2}\partial^{*2}_Z\right) \myexpval{ \psi(\vec{X})\psi(\vec{Y})\psi(\vec{Z})} \; .
\end{align}
Defining $\vec{x} = \vec{X - Z}$ and $\vec{y} = \vec{Y - Z}$, the following equations hold:
\begin{align}
      \myexpval{ \hat{\kappa}\hat{\kappa}\hat{\kappa}} (\vec{\ell_1}, \vec{\ell_2}, \vec{\ell_3}) = {}&{} -\frac{\pi^4}{6\alpha^5}\ell_1^2\ell_2^2\ell_3^2\,\diracd(\vec{\ell_1}+\vec{\ell_2}+\vec{\ell_3})\ee^{-(\ell_1^2+\ell_2^2+\ell_3^2)/12\alpha} \nonumber\\
      = {}&{} -\frac{\pi^4}{6\alpha^5}\ell_1^2\ell_2^2(\ell_1^2+\ell_2^2+2\vec{\ell_1}\cdot\vec{\ell_2})\,\diracd(\vec{\ell_1}+\vec{\ell_2}+\vec{\ell_3}) \ee^{-(\ell_1^2+\ell_2^2+\vec{\ell_1}\cdot\vec{\ell_2})/6\alpha}\\
      = {}&{} (2\pi)^2B_\kappa(\vec{\ell}_1,\vec{\ell}_2,\vec{\ell}_3) \diracd(\vec{\ell_1}+\vec{\ell_2}+\vec{\ell_3})\; ,\\
      \myexpval{ \gamma\gamma\gamma}  (\vec{x},\vec{y}) \overset{(*)}{=} {}&{} \alpha^3 \left[(\vec{x}-2\vec{y})(\vec{x}+\vec{y})(\vec{y}-2\vec{x})\right]^2\, \ee^{-2\alpha(x^2+y^2-\vec{x}\cdot \vec{y})}\; .
      %
      \label{eq:gammagammagamma_analytic_model}
\end{align}
In the last equation, marked by $(*)$, the variables $\vec{x},\vec{y}$ are interpreted as complex numbers $\vec{x}=x_1+\ii x_2$; for their scalar product, $\vec{x}\cdot\vec{y}=x_1y_1+x_2y_2$ holds.
The equation for $\langle\gamma\gamma\gamma^*\rangle$ was computed via \textsc{Mathematica} and is too long to denote here.
We can now set
\begin{equation}
      b(\ell_1,\ell_2,\varphi) = -\frac{\pi^2\ell_1^2\ell_2^2(\ell_1^2+\ell_2^2+2\ell_1\ell_2\cos\varphi)}{24\alpha^5}\ee^{-(\ell_1^2+\ell_2^2+\ell_1\ell_2\cos(\varphi))/6\alpha}\, .
      \label{eq:bkappa_analytic_model}
\end{equation}
We then transform the Cartesian shears in Eq.~\eqref{eq:gammagammagamma_analytic_model} to the orthocenters using Eq.~\eqref{eq:rotation_orthocenter}, and use the fact that $\Gamma^{(0)}=\expval{\gamma\gamma\gamma}$ and $\Gamma^{(3)}=\expval{\gamma\gamma\gamma^*}$ (see Eq.~\ref{eq:defn_natural_components}), to test our integration routine using an analytic model. As can be seen in Fig.~\ref{fig:shear_3pcf_analytic_model}, the integration routine is accurate to the sub-percent level.

\section{Additional figures}
\begin{figure*}[!hbt]
    \centering
    \includegraphics[width=17cm]{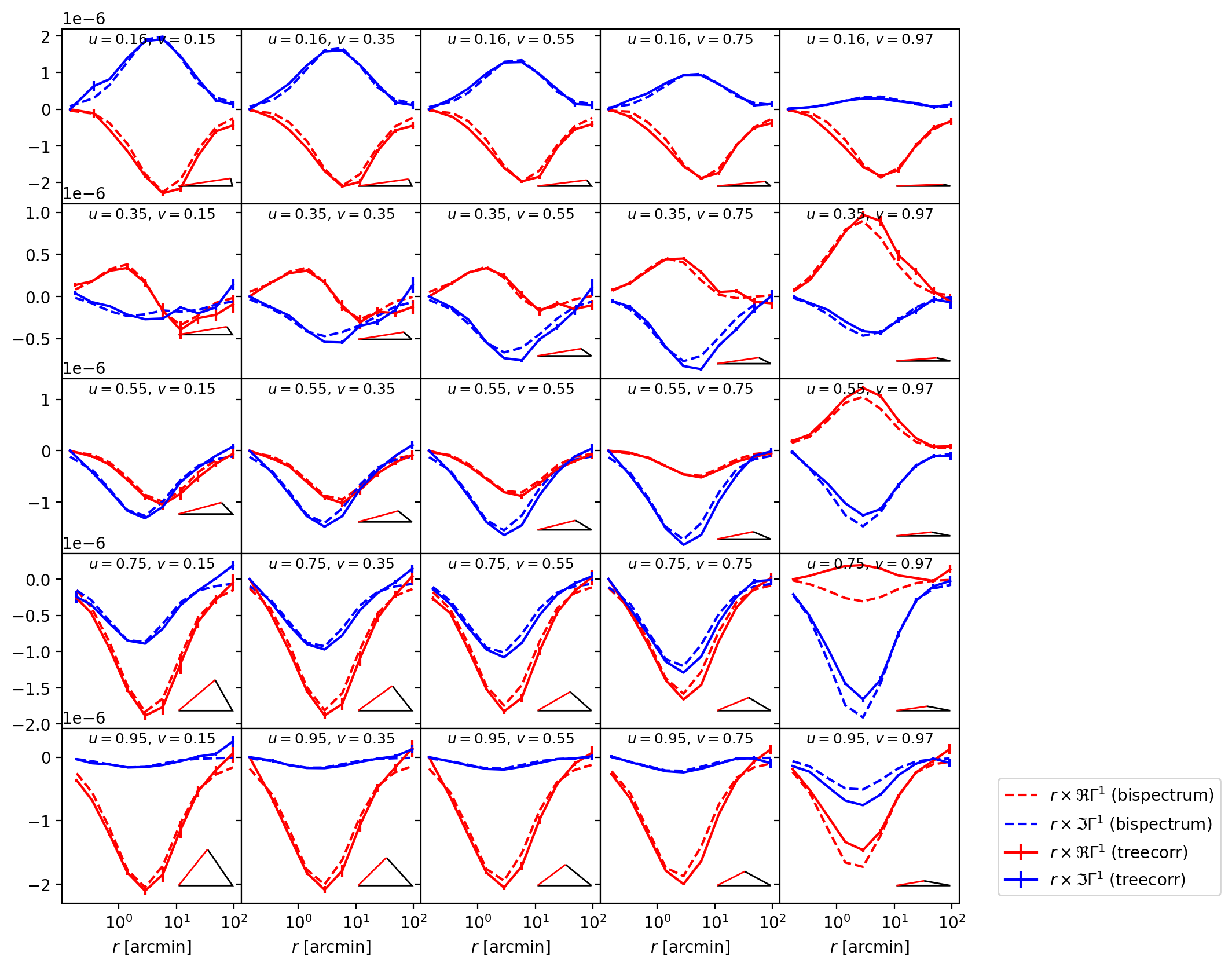}
    \caption{Same as Fig.~\ref{fig:gamma0_bihalofit_vs_MS}, just for the second natural component $\Gamma^{(1)}$.}
    \label{fig:gamma1_bihalofit_vs_MS}
\end{figure*}

\begin{figure}[!hbt]
\centering
\sidecaption
\includegraphics[width=12cm]{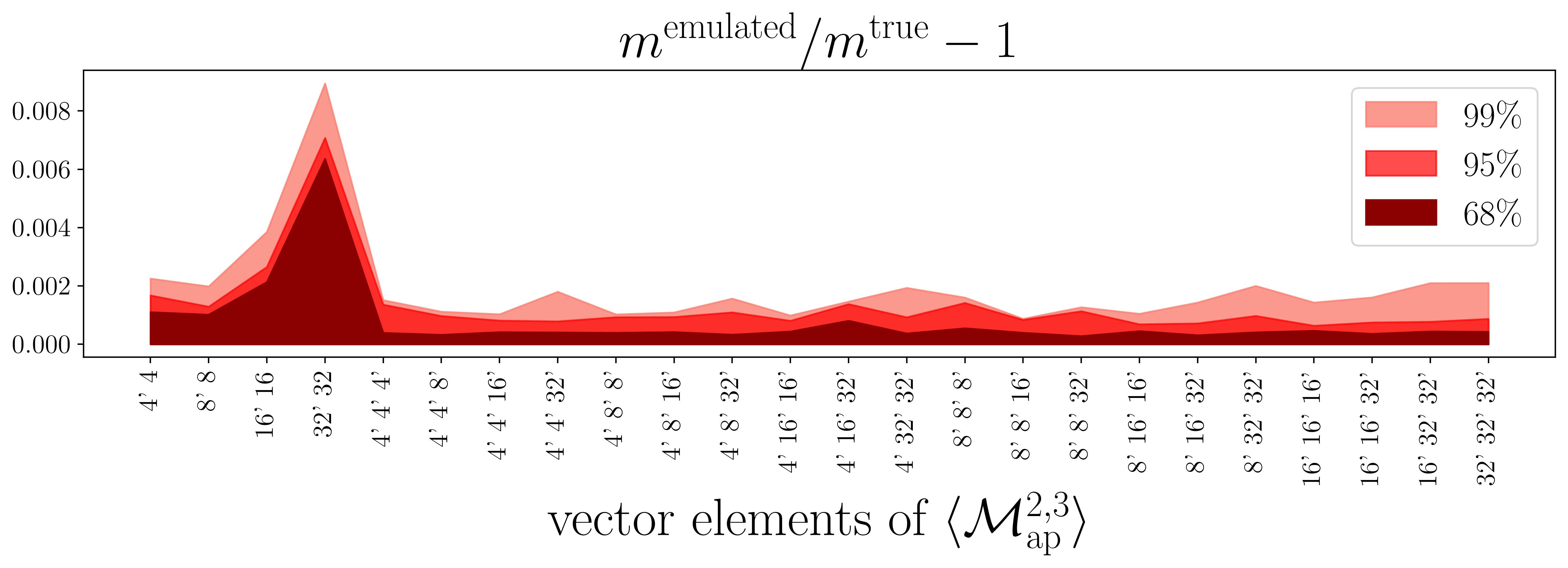}
\caption{Accuracy of the emulated aperture statistics model data vector $m(\boldsymbol{\Theta})$ for the KiDS-1000 setup. Given the bispectrum accuracy of $10\%$ \citep{Takahashi:2020}, the emulator uncertainty plays a negligible role in the modelling process.}
\label{fig:emulator_acc}
\end{figure}

\begin{figure}[!htb]
\centering
\sidecaption
\includegraphics[width=12cm]{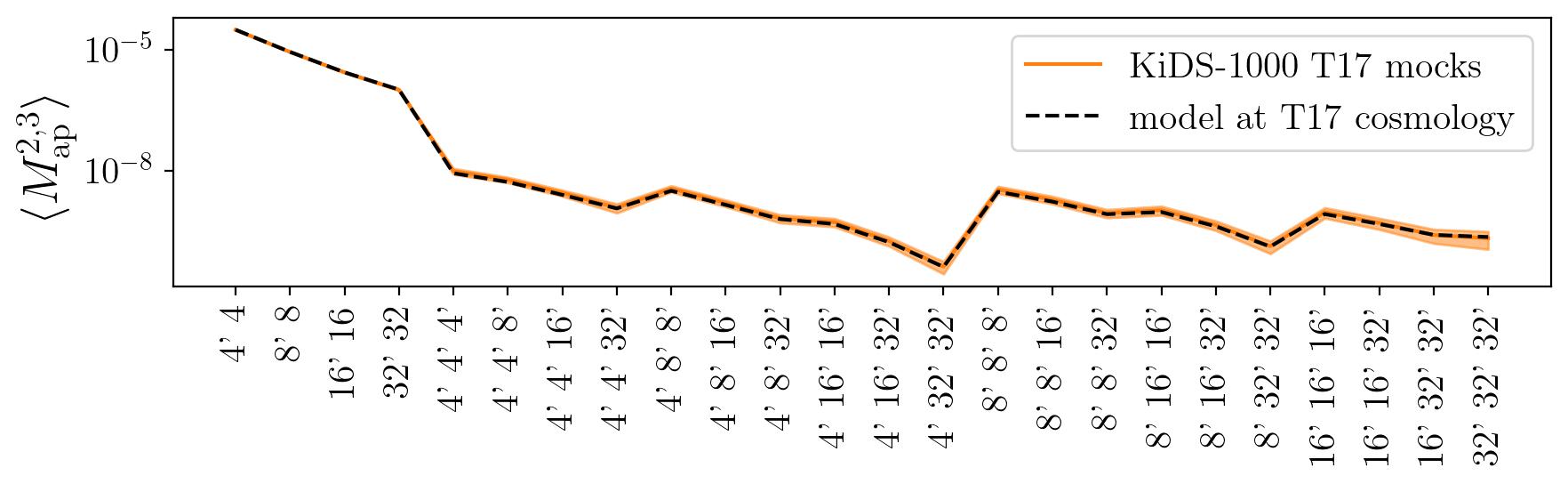}
\caption{Comparison of the measured KiDS-1000-like data vector in the T17 simulations to the modelled vector of the second- and third-order aperture statistics. The orange band is the expected KiDS-1000 uncertainty.}
\label{fig:Map23_vector}
\end{figure}

\end{document}